\PassOptionsToClass{10pt}{revtex4-2}
\documentclass[aps,prx,twoside,twocolumn,superscriptaddress,longbibliography]{revtex4-2}
\usepackage[utf8]{inputenc}
\usepackage{amsmath}
\usepackage{amssymb}
\usepackage{amsfonts}

\usepackage{graphicx}
\usepackage{lipsum}
\usepackage{esint}
\usepackage{times}
\RequirePackage[normalem]{ulem}
\usepackage{physics}
\usepackage{multirow}
\usepackage{hyperref} 

\makeatletter
\usepackage{bm}
\usepackage{xcolor}
\usepackage{dcolumn}
\usepackage[english]{babel}

\usepackage{ulem}

\begin{document}

\title{
Comprehensive Study of Phonon Chirality under Symmetry Constraints
}

\author{Shuai Zhang}
\affiliation{Institute of Theoretical Physics, Chinese Academy of Sciences, Beijing 100190, China}
\author{Zhiheng Huang}
\author{Muchen Du}

\author{Tianping Ying}
\author{Luojun Du}
\affiliation{Institute of Physics, Chinese Academy of Sciences, Beijing 100190, China}
\author{Tiantian Zhang}
\email{ttzhang@itp.ac.cn} 
\affiliation{Institute of Theoretical Physics, Chinese Academy of Sciences, Beijing 100190, China}

\begin{abstract}

Phonons are quanta of lattice vibrations, and their modes (linear, circular, or stationary) are symmetry-determined. Circularly polarized phonons, possessing nonzero angular momentum (AM), have drawn widespread attention recently.
Despite widespread use of pseudo-angular momentum (PAM) and circularly polarized light polarization flips to identify chiral phonons in Raman scattering, their reliability is debated due to symmetry dependence, and experimental verification standards remain lacking.
Here, we systematically study phonon chirality and associated phenomena across magnetic point groups.
We establish that the AM-PAM correlation is governed by both crystalline symmetry and Wyckoff positions, dictating conditions where nonzero AM manifests in PAM signatures. Crucially, phonons belonging to distinct irreducible representations exhibit distinct experimental benchmarks, enabling direct determination of crystalline chirality and symmetry classification. Furthermore, we report the discovery of a signature for symmetry-induced phenomena, notably a half-wave plate-analogous effect induced by mirror-odd phonons.
Meanwhile, we conducted five experiments to validate our theory.

\end{abstract}

\maketitle

\section{Introduction}

Phonons carrying nonzero angular momentum (AM), historically termed ``circularly polarized phonons'' or ``rotational vibrations''~\cite{johnson1982angular,rebane1983faraday,mclellan1988angular}, are now predominantly called chiral phonons~\cite{bermudez_chirality_2008,bermudez2008hyper, Niu_phonon_AM_2014, Te_rotation_2018}.
Recent advances underscore the significance of chiral phonons as collective excitations with mechanical rotations~\cite{ Sinisa_PT_AM_2023, Te_rotation_2018,zhang2024observation}. Their interactions with other quasiparticles can produce novel phenomena, most notably giant phonon magnetism observed in diverse materials~\cite{ Niu2021_phononMag, Niu2022_MolecularBerry, Xue2025_Extrinsic, Ren_PH_SPIN_PRX_2024, FeGeTe_2D_chiral_phonon_Du_2019, CeCl3_2022, Bonini2023_CrI3_TRS, Liu2021_Magnon_phonon, CoSnS_CP_2025, CoSnS_PRL_2025}. This magnetism arises from ionic circular motion and is explained by molecular Berry curvature (MBC)~\cite{Nagaosa2010_THE,Tao2012_Berry_THE, Nagaosa2019_Berry_THE,
Niu2022_MolecularBerry, Xue2025_Extrinsic} originating from electron-phonon coupling. MBC underpins intriguing phenomena, including the thermal Hall effect~\cite{Strohm2005_THE, Nagaosa2010_THE_QuantumMagnets,
Nagaosa2010_THE, Ideue2017_THE_multiferroics, Zhang2021_ATHE_VI3}, further linking it to chiral phonons.

Currently, pseudo-angular momentum (PAM), defined by the eigenvalues of $C_n$\cite{PAM_Niu2008, Niu_Chiral_phonon_2015, Zhang_2022_screw_PAM}, serves as an alternative identifier for chiral phonons through indirect probes such as circular polarized Raman scattering (CPRS)~\cite{ishito2023truly, Zhang2023_Te, Marc2025_CP,CoSnS_CP_2025} and infrared spectroscopy~\cite{zhu2018observation,CoSnS_PRL_2025}. This approach circumvents experimental challenges in direct AM detection of phonons~\cite{MoO3_CP_2024,zhu2018observation,ishito2023truly, Zhang2023_Te, Ishito2023_Te, Marc2025_CP,zhang2025weyl,zhang2025new}, yet three fundamental questions persist:
(i) Does nonzero PAM imply nonzero AM?
(ii) Are phonons in cross-circular scattering truly chiral?
(iii) What criteria experimentally confirm chiral phonons?

\begin{figure*}
  \centering
  \includegraphics[width=1.0\textwidth]{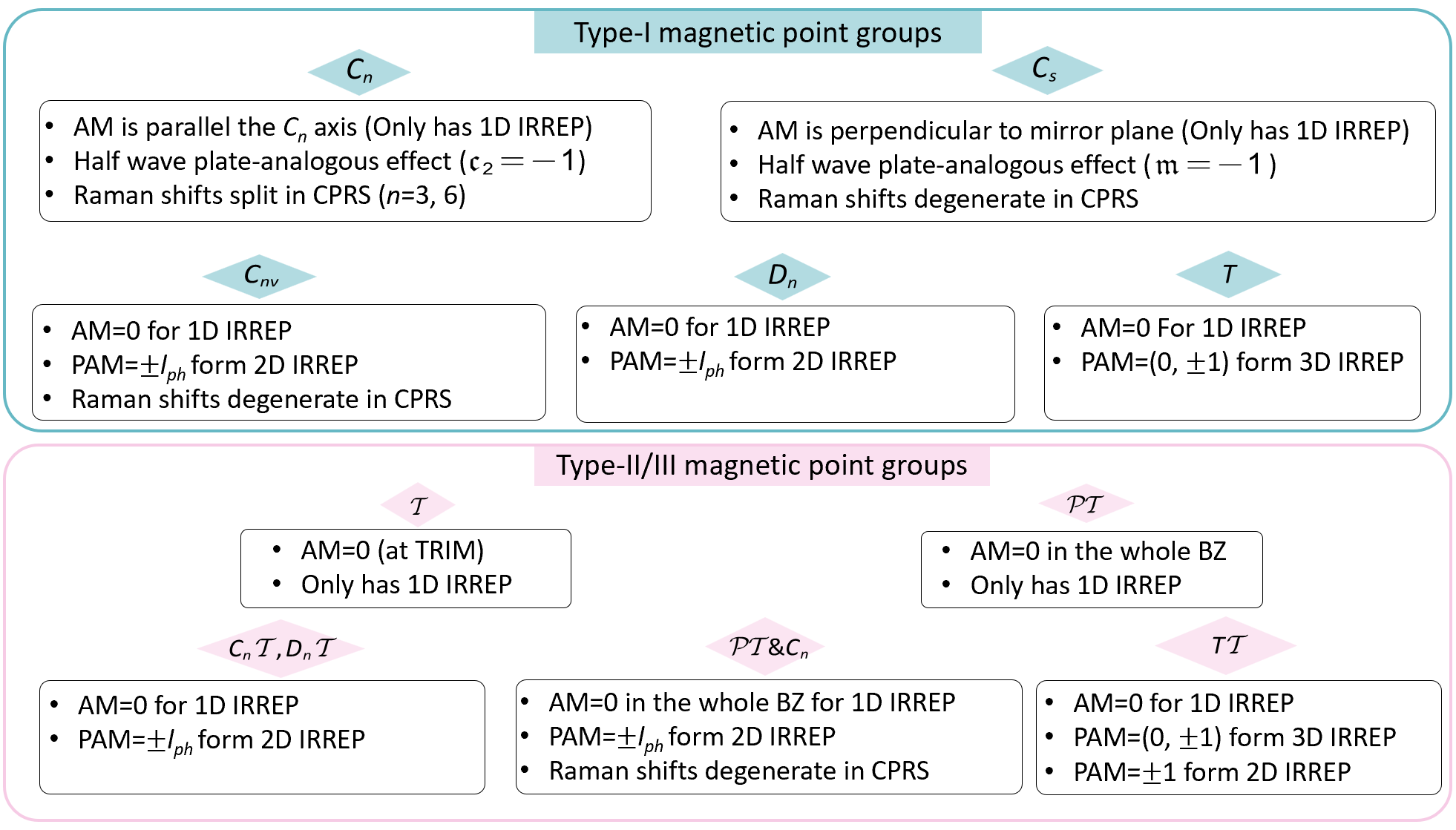}
  \caption{\label{fig:chiral_phonon_symmetry_flow}
Summary of the results on phonon chirality under representative symmetries, crossing type-I, type-II, and type-III magnetic point groups. Symmetries shown in the diamond box represent the little group of momenta $\boldsymbol {q}$ in the Brillouin zone (BZ).
$\mathcal{P}$ and $\mathcal{T}$ denote inversion and time-reversal symmetry, respectively.
{$\mathfrak{c}_2$ and $\mathfrak{m}$ represent the eigenvalue of the twofold rotation ($C_2$) and the mirror ($\mathcal{M}$) operation in the corresponding point groups, ``1D IRREP'' is the abbreviation of ``one-dimensional irreducible representation''}. Results for other point groups not shown here can be deduced from these representative cases, with detailed discussions provided in the corresponding sections of the main text. The AM of non-degenerate phonon modes is constrained to zero by point group symmetries, while the AM of modes in high-dimensional IRREPs may range from $-1$ to $1$ and cannot be determined simultaneously due to superposition.
}
\end{figure*}

Since symmetry governs scattering processes (e.g., electron-, light-, magnon-phonon), we systematically investigated phonon chirality across type I, II, and III magnetic point groups.
We find that the relationship between AM, PAM, and cross-circular polarized Raman scattering critically depends on symmetries and occupied Wyckoff positions. These results, including the novel \textit{half-wave plate}-analogous effect and symmetry conditions enabling AM-PAM correlation, are summarized in Figure~\ref{fig:chiral_phonon_symmetry_flow} and Tables S2, S3, S4, and S5.

This work is organized as follows: we begin in Section~\ref{sec:chiral_phonon_definition} by revisiting key concepts related to chiral phonons and introducing some nomenclature. Subsequently, we 
investigate the chirality of phonons under various symmetries. We begin with type-I magnetic group symmetries, such as rotation symmetries $C_{n}$ in Section~\ref{sec:chiral_phonon_under_rotation}; then discuss the relationship between AM and PAM in Section~\ref{sec:AMPAM};
mirror symmetry ($\mathcal{M}$) in Section~\ref{sec:mirror}, the combination of mirror and rotation symmetry ($C_{nv}$ {and} $C_{nh}$ point groups) in Section~\ref{sec:mirror_rotation}; point groups composed of multiple rotation axes in Section~\ref{sec:Dn_group} ($D_n$, {$D_{nd}, D_{nh}$}) and Section~\ref{sec:TO_group} ($T, T_d, T_h$ and $O, O_h$);
type-II magnetic point group with {time-reversal symmetry {($\mathcal{T}$)}} in Section~\ref{sec:time_reversal_symmetry}, type-II/III magnetic point group with $\mathcal{PT}$ symmetry in Section~\ref{sec:inversion_chiral}. 
The main results of the comprehensive study are summarized in Fig.~\ref{fig:chiral_phonon_symmetry_flow}.
Last but not least, we conduct experiments on five materials with distinct symmetries to validate our theoretical results in Section~\ref{sec:exp}.
We will make a brief conclusion and discussion in Section~\ref{sec:conclusion}.

\section{Results}

\subsection{Revisiting phonon chirality fundamentals}

\label{sec:chiral_phonon_definition}

Phonon AM for a specific mode can be expressed as the polarization vector, i.e., the eigenvector of the dynamic matrix~\cite{johnson1982angular,rebane1983faraday,mclellan1988angular,Niu_phonon_AM_2014}.
Under the harmonic approximation, phonon mode $\boldsymbol {\epsilon}_{\nu \boldsymbol {q}}$ is the eigenvector of the mass-weighted dynamic matrix {$D_{\kappa\kappa'}^{\alpha\beta}(\boldsymbol {q})$}, i.e., 
\begin{equation}
    \begin{aligned}
 \sum_{\beta \kappa'} D^{\alpha \beta} _{\kappa\kappa'}  (\boldsymbol {q}) \boldsymbol {\epsilon}^{\beta\kappa'}_{\nu\boldsymbol {q}} = \omega^2_{\nu\boldsymbol {q}}\boldsymbol {\epsilon}^{\alpha\kappa}_{\nu \boldsymbol {q}}.
    \end{aligned}
\end{equation}
$\alpha/\beta \in \{x, y, z\}$ and $\kappa/\kappa'$ represents the index of atoms in the primitive cell.
$\omega_{\nu\boldsymbol {q}}$ is the frequency of the $\nu$-th mode at $\boldsymbol {q}$. 
$\epsilon_{\nu\boldsymbol {q}}$ can be expressed in the complex amplitude form of
\begin{equation}
\begin{aligned}
    \label{eq:wf_compl_amp}
    \epsilon_{\nu\boldsymbol {q}} & =\bigoplus_\kappa \epsilon^\kappa_{\nu\boldsymbol{q}}, \text{and} \\
    \boldsymbol {\epsilon}^\kappa_{\nu\boldsymbol {q}} & =\{ A^{\kappa}_{x} e^{i\theta^{\kappa}_{x}}, A^{\kappa}_{y} e^{i\theta^\kappa_{y}}, A^{\kappa}_{z} e^{i\theta^\kappa_{z}}\},
\end{aligned}
\end{equation}
where $A^\kappa_{\alpha}$ and $\theta^\kappa_{\alpha}$ are real numbers.
Hereto, one can define phonon AM for a specific phonon mode $\boldsymbol{\epsilon}_{\nu\boldsymbol{q}}$ at momentum $\boldsymbol{q}$ as~\cite{mclellan1988angular, Niu_phonon_AM_2014}:
\begin{equation}
\label{eq:AM_defination}
\begin{aligned}
  l_{\alpha, \nu\boldsymbol{q}}  &= \hbar \boldsymbol{\epsilon}_{\nu\boldsymbol{q}}^\dagger  M_{\alpha}  \boldsymbol{\epsilon}_{\nu\boldsymbol{q}} \\
  & = \sum_{\kappa}^{N} l^\kappa_{\alpha, \nu\boldsymbol{q}}=\sum_{\kappa}^{N} \hbar \boldsymbol{\epsilon}^{\kappa\dagger}_{\nu\boldsymbol{q}} \mathfrak{M}_{\alpha} \boldsymbol{\epsilon}^{\kappa}_{\nu\boldsymbol{q}}.
  \end{aligned}
\end{equation}
$M_{\alpha} = \oplus_{\kappa=1}^{N} \mathfrak{M}_{\alpha}$, where $\mathfrak{M}_{\alpha(\beta\gamma)}=(-i) \varepsilon_{\alpha(\beta\gamma)}$ forms the Lie algebra of the $SO(3)$ group, {$\varepsilon_{\alpha(\beta\gamma)}$ is the Levi-Civita tensor,} and $N$ is the number of atoms in the  primitive cell.
The $\alpha$-component of AM for the $\kappa$-th atom, can {also} be expressed by the phase difference between the $\beta$ and $\gamma$ component, i.e., 
\begin{equation}
\label{eq:relative_phase}
    l^\kappa_{\alpha, \nu\boldsymbol {q}}= 2\text{Im}[A^{\kappa}_\beta A^{\kappa}_\gamma e^{i\epsilon_{\alpha(\beta\gamma)} (\theta^\kappa_\beta-\theta^\kappa_\gamma)}].
\end{equation}
A detailed proof is shown in Supporting Information Section S1.

We note phonon modes with zero AM do not necessarily exhibit linear atomic vibrations, counter-rotating atoms with opposite circular polarization is also possible.
We also note that $l_{\alpha, \nu\boldsymbol {q}}$ is always real, and $\boldsymbol {l}_{\nu\boldsymbol {q}}$ constitutes a pseudo-vector field in Brillouin zone (BZ), {transformed} as a vector under $SO(3)$ rotations but remaining invariant under inversion operation ({$\mathcal{P}$)}. The chirality of the phonons defined by AM can change under different reference coordinates owing to the pseudovector nature.
Thus, there also exists an alternative definition of chiral phonon based on the nonzero \textit{helicity}~\cite{ishito2023truly, Parlak2023_ph_helicity, CP_helicity_APL_2024, HelicityMBC_2021}, expressed as:
\begin{equation}
\label{eq:helicity}
    h_{\nu\boldsymbol {q}}=\boldsymbol {q}\cdot\boldsymbol {l}_{\nu\boldsymbol {q}}.
\end{equation}
The helicity is a pseudo-scaler and remains invariant under $SO(3)$ operations, i.e., it is convention-independent and well-defined. Additionally, this definition can be linked to the state of chiral charge density waves (CCDW)~\cite{TA_CDW_Dai2023, zhang2024_CCDW, Romao2024_phonon_induces_geometric_chirality,yang2022visualization, Mauro2024, Eric2024}. In this work, we focus on the widely used definition of chiral phonons based on AM, i.e., the circularly polarized phonons~\cite{rebane1983faraday,mclellan1988angular,bermudez_chirality_2008,bermudez2008hyper,Niu_phonon_AM_2014, Te_rotation_2018}.
Extended discussions on phonon helicity and its relationship to CCDW are provided in the Supporting Information  Section S2.

In the following sections, we systematically explore chiral-phonon properties across different magnetic little groups of $\boldsymbol {q}$, addressing the main text's core questions.

\subsection{$C_n$ Rotation Symmetry}
\label{sec:chiral_phonon_under_rotation}
In this section, we study the chirality of phonon at momentum $\boldsymbol {q}$ where the little group has only $C_n$ symmetry, which corresponds an Abelian group only have one-dimensional {(1D)} {irreducible representations (IRREPs)}.

\subsubsection{AM under $C_n$}
\label{sec:rotation_on_AM}
As we mentioned {ealier}, $\boldsymbol {l}_{\nu\boldsymbol {q}}$ forms a pseudo-vector field in BZ under $O(3)$ operations. Thus, 

{\textit{if momentum $\boldsymbol {q}$ has $C_n$ symmetry,  $\boldsymbol {l}_{\nu\boldsymbol {q}}$ must parallel to the rotation axis.}}

\subsubsection{PAM and CPRS under $C_n$}
\label{sec:CnRaman}

\begin{figure}[h]
  \centering
  \includegraphics[width=0.5\textwidth]{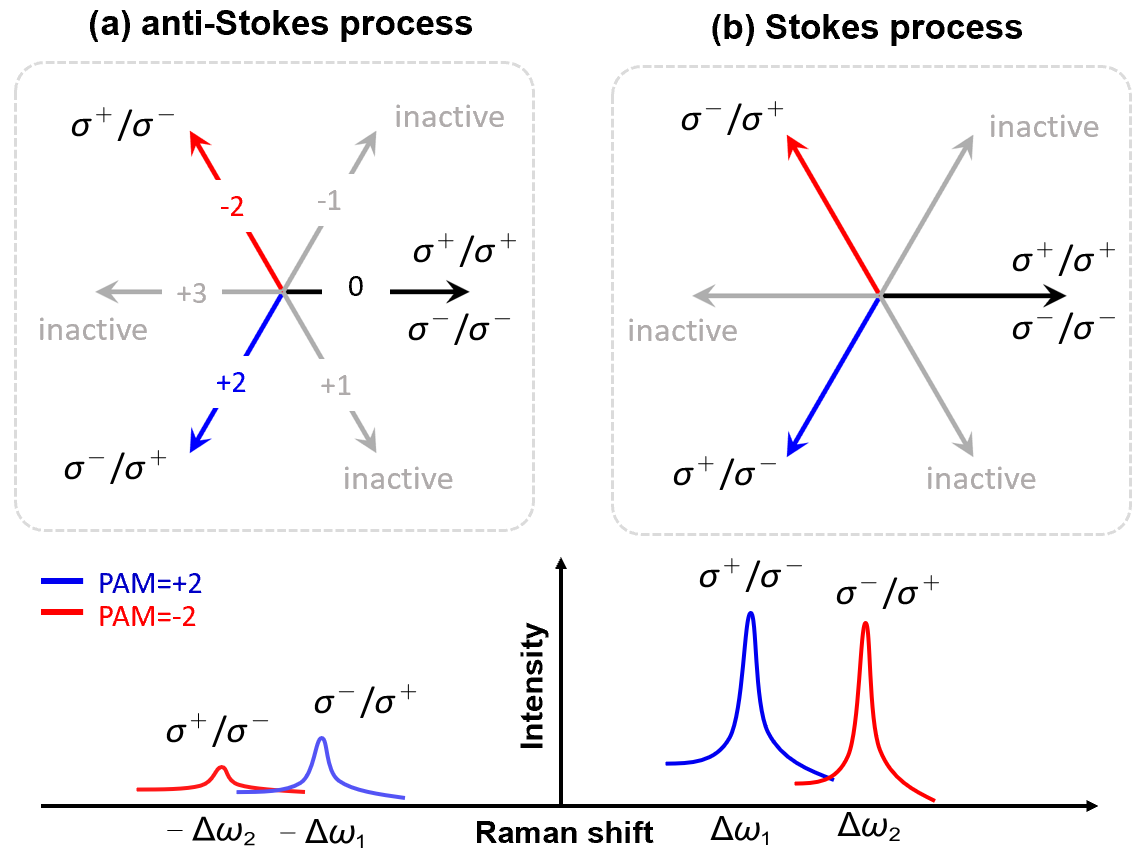}
  \caption{\label{fig:Raman_antiRaman} The selection rules for the CPRS within a $C_6$-symmetric system, assuming the light propagation along the $C_6$ axis. The arrows symbolize the eigenvectors of $C_6$ operation, marked by the value of PAM$=0,\pm1,\pm2,3$, with the permissible CPRS explicitly indicated. Phonon modes with different PAMs are active under different CPL settings and in different Raman processes. We note that the phonon mode with PAM=$-2$ is active only in the $\sigma^-/\sigma^+$ Stokes process. However, according to the Raman tensor, it is active for both the Stokes and anti-Stokes processes.}
\end{figure}

The pseudo-angular momentum (PAM{, $l_{ph}$}) is defined through the eigenvalues of the representation matrix of $C_n$, i.e.,
\begin{equation}
    \label{eq:PAM_defination}
    \mathcal{D}(C_n) {\epsilon}_{\nu\boldsymbol {q}}=e^{{-}i 2\pi l_{ph}/n} {\epsilon}_{\nu\boldsymbol {q}}. 
\end{equation}
PAM has been widely discussed with chiral phonons by CPRS in $C_3$-invariant systems~\cite{PAM_Niu2008,miao2018observation, Niu_Chiral_phonon_2015, zhu2018observation,li2021observation,  Zhang_2022_screw_PAM, Zhang2023_Te, ishito2023truly, David_2024, Thuc_CoTiO_2024, CoSnS_PRL_2025,zhang2025weyl,zhang2025new}.
Apart from this, we systematically studied the {CPRS} selection rules in $C_{n=2,3,4,6}$-invariant systems for both Stokes/anti-Stokes processes.
We assume that incident/scattered circularly polarized light (CPL) propagates along the $C_n$ axis~(assumed to be $z+$). 
This request that the phonon momentum $\boldsymbol{q}$ on the $C_n$ axis path through the $\boldsymbol{\Gamma}$ point (i.e, $C_n\boldsymbol{q}=\boldsymbol{q}$) for the momentum conservation; otherwise the propagating direction of the scattering light will be off the $C_n$ axis and CPL can not be the eigenstate of $C_n$ (See Supporting Information Section S3
for details).
The polarization of CPL includes the right hand ($\sigma^+=(1, i, 0)^T$, PAM=$+1$ under $C_{n=3,4,6}$ and PAM={$1$} under $C_{n=2}$) and left hand ($\sigma^-=(1, -i, 0)^T$, PAM=$-1$ under $C_{n=3,4,6}$ and PAM={$1$} under $C_{n=2}$). There are four possible CPRS {processes} based on the combinations of the incident/scattered CPL, i.e., $\sigma^+/\sigma^+$,$\sigma^-/\sigma^-$, $\sigma^+/\sigma^-$, and $\sigma^-/\sigma^+$. 
The exhaustive results are shown in the Supporting Information  Section S4.

We note that although the Raman scattering tensor for specific phonon modes encodes all the allowed scattering processes, it can not distinguish the Stoke/anti-Stokes (phonon-emission/absorption) processes.
However, the CPRS selection rules for these two processes are different.
Taking the $C_6$ invariant condition as an example, as shown in Figure~\ref{fig:Raman_antiRaman}. The phonon mode with PAM=$-2$ is only active in the $\sigma^-/\sigma^+$ Stokes process, and it is inactive in the anti-Stokes process. However, according to the Raman tensor, the phonon mode with PAM=$-2$ is active for both of the Stokes/anti-Stokes processes. Thus, the selection rule addresses Raman tensor limitations in CPRS experiments.
Detailed discussions are in the Supporting Information Section S5. 

\subsection{AM-PAM relationship: Symmetry and Wyckoff Position Dependence}
\label{sec:AMPAM}
Although nonzero PAM is commonly used to identify chiral phonons~\cite{PAM_Niu2008,miao2018observation, Niu_Chiral_phonon_2015, zhu2018observation,li2021observation,  Zhang_2022_screw_PAM, Zhang2023_Te, ishito2023truly, David_2024, Thuc_CoTiO_2024, CoSnS_PRL_2025,zhang2025weyl,zhang2025new}, we emphasize that this criterion is not universal: its validity depends critically on both the system's symmetry and the occupied Wyckoff positions.
PAM is rotation-center independent at-invariant q, where most PAM-conserving scattering occurs.

\subsubsection{General Wyckoff Positions}
\label{sec:PAM_AM_relationship}

For atoms at general Wyckoff positions {(i.e., the identity site symmetries)}
under $C_n$ symmetry (along $z$), Schur decomposition of $\mathcal{D}(C_n)$ yields orthonormal eigenvectors $\epsilon_i$ with defined PAM and AM ($l_{z,i}$), and the AM of the $\epsilon_i$ with the same PAM can have opposite sign, we illustrate this in Supporting Information Section S6 
with a tight-binding model.
For any {eigenstates of $D(\boldsymbol{q}$}) (denoted as $\epsilon_{\nu\boldsymbol{q}}$) with $C_n\boldsymbol{q} = \boldsymbol{q}$, it decomposes into $\epsilon_i$ sharing its PAM. Suppose in $C_3$-symmetric systems, $\epsilon_1$ and $\epsilon_2$ both have PAM = $+1$ but opposite-sign AM, $\epsilon_{\nu\boldsymbol{q}}$ with PAM=$+1$ can be expressed as:
\begin{equation}
    \epsilon_{\nu\boldsymbol{q}}=a\epsilon_1+b\epsilon_2,
\end{equation}
where $a, b\in \mathbb{C}$, and $|a|^2+|b|^2=1$.
The AM of $\epsilon_{\nu\boldsymbol{q}}$ can be expressed as:
\begin{equation}
\begin{aligned}
    l_{z, \nu\boldsymbol{q}}&=\langle\epsilon_{\nu \boldsymbol{q}}|M_z|\epsilon_{\nu \boldsymbol{q}}\rangle\\
    &=|a|^2 l_{z, 1} + |b|^2 l_{z, 2} + 2\text{Re}[a^*b\langle\epsilon_1|M_z|\epsilon_2\rangle].
\end{aligned}
\end{equation}
By tuning the force constants-related parameters $a$ and $b$, while preserving $C_3$ symmetry, $l_{z,\nu\boldsymbol{q}}$ can continuously vary from $-1$ to $+1$ (including 0). Thus, at general Wyckoff positions, phonons with nonzero PAM may exhibit zero AM.
{Moreover,} if there are additional symmetries, AM can be restricted to zero for the phonon modes with nonzero PAM. An example involves the phonon modes with PAM $= 2$ under $C_{4v}$ symmetry exhibit zero AM. This behavior arises due to the constraints imposed by vertical mirror symmetry. A detailed discussion of this phenomenon and experimental observation will be presented in Section~\ref{sec:cnv_2d} and Section~\ref{sec:exp_C4}. 
These result indicates that there is no intrinsic relationship between the PAM and AM in the general cases.

\subsubsection{$C_n$-symmetric Wyckoff Positions}
\label{sec:AM_Cn_Wyckoff}

When atoms are at occupied $C_n$-symmetric Wyckoff positions, the AM-PAM relationship at $\boldsymbol{q}$ that fulfills $C_n\boldsymbol{q}=\boldsymbol{q}$ can be established by symmetry. 
In this case, the representation matrix of $C_n$ reads (details are in Supporting Information Section S7 
):
\begin{equation}
\mathcal{D}(C_n)= \mathbb{I}_{N \times N} \otimes C_n,
\end{equation}
where $\mathbb{I}_{N \times N}$ is the identity matrix, and $C_n$ is the Euclidean representation matrix. With the $C_n$ axis along $z$, only $z$-component AM can be nonzero. For $n=3,4,6$ and atoms at $C_n$-invariant Wyckoff positions, only PAM=$0,\pm1$ {modes} arise at $\boldsymbol{q}$.
The eigenvectors of {$\mathcal{D}(C_n)$ reads:}
\begin{equation}
\begin{aligned}
    \epsilon_{i} &= (0,0,0,...,\epsilon^{\kappa}_{i},...,0,0,0)^T \text{ with}\\
    \epsilon^{\kappa}_{i} &= \frac{1}{\sqrt{2}}(1, \pm i, 0),
\end{aligned}
\end{equation}
which have $l_{ph, i}=\pm 1$ and ${l_{z,i}}=\pm 1$.
Meanwhile, phonon modes with eigenvectors of
\begin{equation}
\begin{aligned}
    \epsilon_{\nu \boldsymbol{q}} &= (0,0,0,...,\epsilon^{\kappa}_{\nu \boldsymbol{q}},...,0,0,0)^T \text{and}\\
    \epsilon^{\kappa}_{\nu \boldsymbol{q}} &= (0, 0, 1)
\end{aligned}
\end{equation}
have $l_{ph, i}=0$ and ${l_{z,i}}=0$.
Any eigenvector of \( D(\boldsymbol{q}) \) with PAM=+1 decomposes into basis states \(\epsilon_i\) sharing PAM=+1 and \(l_{z,i}=+1\):
\[
\epsilon_{\nu\boldsymbol{q}} = \sum_{i=1}^{n_{+1}} a_i \epsilon_i \text{and} \quad a_i \in \mathbb{C}, \quad \sum_{i=1}^{n_{+1}} |a_i|^2 = 1,
\]
where \(n_{+1}\) counts \(\mathcal{D}(C_n)\) eigenvectors with PAM=+1. The angular momentum \(l_{z,\nu\boldsymbol{q}}\) then follows:

\begin{equation}
\begin{aligned}
    l_{z, \nu\boldsymbol{q}}&=\langle\epsilon_{\nu \boldsymbol{q}}|M_z|\epsilon_{\nu \boldsymbol{q}}\rangle\\
    &=\sum_{i=1}^{n(+1)} |a_i|^2\cdot l_{z, i} + \sum_{i<j}^{n(+1)} 2\text{Re}[a_i^*a_j\langle\epsilon_i|M_z|\epsilon_j\rangle].
\end{aligned}
\end{equation}
Crucially, each $\epsilon_i$ simultaneously diagonalizes $\mathcal{D}(C_n)$ and $M_z$ (eigenvalue +1 for $M_z$), yielding $\langle\epsilon_i| M_z |\epsilon_j\rangle = \langle\epsilon_i|\epsilon_j\rangle = \delta_{ij}$. Consequently, phonon modes with PAM=\(+1\) will have $l_{z,\nu\boldsymbol{q}} = +1$. {Likely,} phonon modes with PAM=$-1$ will have $l_{z,\nu\boldsymbol{q}} = -1$ and phonon modes with PAM=\(0\) have $l_{z,\nu\boldsymbol{q}} = 0$.

For $C_4$-invariant systems with atoms at $C_2$-invariant Wyckoff position, the AM ($l_{z,\nu\boldsymbol{q}}$) of phonon modes with PAM=$0, \pm1$ can not be determined, while $l_{z,\nu\boldsymbol{q}}=0$ for phonon modes with PAM=$2$ since they are pure $z$-polarized.
In $C_6$-invariant systems, if the atoms occupy the $C_3$-symmetric Wyckoff, $l_{z,\nu\boldsymbol{q}}=+1$ when PAM $\in {-2,+1}$; $l_{z,\nu\boldsymbol{q}}=-1$ when PAM $\in {+2,-1}$; and $l_{z,\nu\boldsymbol{q}}=0$ (purely $z$-polarized) when PAM $\in {0,3}$. If the atoms occupy the $C_2$ -symmetric Wyckoff $3c$, there is no AM-PAM relationship.
In $C_2$-invariant systems, the sole definitive AM-PAM relationship emerges when atoms occupy the $C_2$-invariant Wyckoff position: phonon modes with PAM $=0$ exhibit zero AM ($l_{z,\nu\boldsymbol{q}} = 0$) as they are purely $z$-polarized modes.
These results are summarized in Tables~S2--S5.

\subsection{Mirror symmetry}
\label{sec:mirror}

In this section, we examine chiral phonon-related quantities and phenomena under the little group of $\boldsymbol {q}$ {only} preserves mirror symmetry $\mathcal{M}$ (i.e., the Abelian group $C_s$, {which only has 1D IRREPs}), we have the following conclusion:

\textit{The AM of a non-degenerate phonon is oriented perpendicular to the mirror plane.}

The atomic motion of the mirror-constrained phonon modes depends on the Wyckoff positions. Details of the proof are shown in the Supporting Information Section S8.

\subsubsection{\textit{Half-wave plate}-analogous effect under $\mathcal{M}$} 
\label{sec:half_wave_plate}

\begin{figure*}
  \centering
  \includegraphics[width=0.9\textwidth]{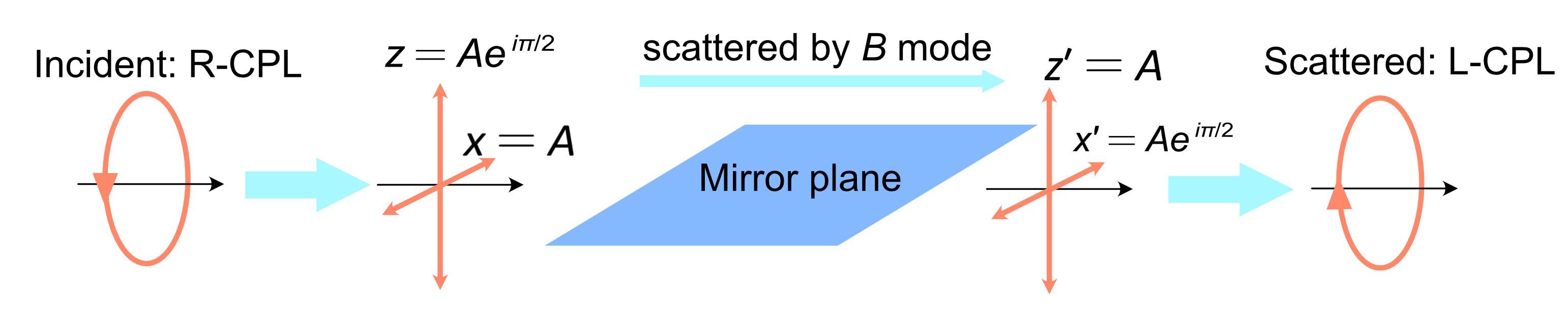}
  \caption{Illustration for the half-wave plate-analogous effect for the phonon mode $B$ in the system with $C_s$ (or $C_{n=2,4,6}$) point group. $B$ is the phonon mode with $\mathfrak{m}_z=-1$. In the Raman scattering process, we set the incident/scattered light propagates along the $y$ axis, which is parallel to the mirror plane.
  If the incident light is right-handed circular polarized~(R-CPL) and scattered by the phonon mode $B$, the scattered light will become left-hand circular polarized light~(L-CPL).}
  \label{fig:mirror_Raman}
\end{figure*}

In this section, we show cross-circular polarization scattering requires neither nonzero PAM nor AM. In systems with a single mirror symmetry, where PAM can not be defined, such scattering occurs via phonons with odd mirror eigenvalues ($\mathfrak{m} = -1$) labeled by $B$ modes.
We discovered that when the CPL propagates parallel to a mirror plane, $B$ modes act as a \textit{half-wave plate}, which will invert the polarization of the light, {meanwhile}, the scattering channel with the same CPL is forbidden. An intuitive picture is shown in Figure~\ref{fig:mirror_Raman}, based on phonon absorption in systems with a single mirror plane $\mathcal{M}_z$. 
This conclusion applies to both Stokes and anti-Stokes processes and is consistent with Raman tensor analysis. 
Details are in the Supporting Information Section S9.

This conclusion generalizes to systems with orthogonal mirror planes. Experimentally validated in Section \ref{sec:exp_Cs}, and we demonstrate that even achiral phonons can invert CPL polarization while forbidding same-circular-polarization scattering channels.

We note that a similar \textit{half-wave plate}-like effect can also {happen} when light propagates perpendicular to the $C_{n=2,4,6}$ axis after scattering with phonon modes with $C_2$ eigenvalue ($\mathfrak{c}_2$) of $-1$. This result is also consistent with the analysis of the Raman tensor.

\subsection{${C}_{nh}$ and ${C}_{nv}$ point group}
\label{sec:mirror_rotation}

While previous sections examined wave vectors $\boldsymbol {q}$ preserving either $C_n$ or $\mathcal{M}$ symmetry, we now consider $\boldsymbol {q}$ points that simultaneously exhibit both symmetries, i.e., $C_{nh}$, in which the $C_n$ axis is perpendicular to the mirror plane, and $C_{nv}$, in which the $C_n$ axis is parallel to the mirror plane.

\subsubsection{$C_{nh}$ point group}
\label{Sec:Cnh}

For $C_{nh}$ point groups, AM constraints inherit rules from both $C_n$- and $\mathcal{M}$-symmetric systems. Crucially, AM orientation remains compatible at $\boldsymbol {q}$ belongs to $C_{nh}$: when parallel to the $C_n$ axis, it automatically lies perpendicular to the horizontal mirror plane $\mathcal{M}$.
For systems with $n$ = 2, 4, 6, $C_{nh}$ symmetry includes inversion ($\mathcal{P}$). Since AM is a pseudo-vector and $\mathcal{P}$ doesn't change its direction, we'll discuss phonon chirality considering both $\mathcal{P}$ and $\mathcal{T}$ in Section~\ref{sec:inversion_chiral}.

\subsubsection{$C_{nv}$ {point group}}

\label{sec:cnv_2d}

If the horizontal mirror {plane} shifts to the vertical one, the little point group for the momentum $\boldsymbol {q}$ becomes $C_{nv}$.
Since the direction of AM under $C_n$ and the vertical mirror $\mathcal{M}$ is incompatible, 
we conclude: 

\textit{The AM of the non-degenerate phonon at $\boldsymbol {q}$ should be zero under little group of $C_{nv}$.}

The above result yields a key insight for $C_{4v}$ symmetry: a phonon with PAM=2 is non-degenerate and thus carries zero AM. This demonstrates that nonzero PAM phonon modes can exhibit vanishing AM.

Unlike the $C_{nh}$ point groups, which only has 1D IRREPs, phonon modes with PAM=$\pm l_{ph}$ will be degenerate under the little group of $C_{nv} (n=3,4,6$) {and form a two-dimensional (2D) IRREPs}. In the subspace spanned by these states, the phonon AM may range from $-1$ to $+1$, and its value cannot be determined simultaneously due to superposition. However, during specific scattering processes, the phonon AM of the excited mode can be fixed by external stimuli, potentially breaking $\mathcal{M}_{\parallel}$ according to the result of Section~\ref{sec:AM_Cn_Wyckoff}.

The discussion on the AM of phonon modes with the higher-dimensional IRREPs applies to all the magnetic point groups. An example is illustrated and discussed in the Supporting Information Section S12.

\subsubsection{Distinguishing $C_n$ and $C_{nv}$ via CPRS ($n=3,4,6$)}
\label{sec:cnv_3d}

\begin{figure*}
  \centering
  \includegraphics[width=0.97\textwidth]{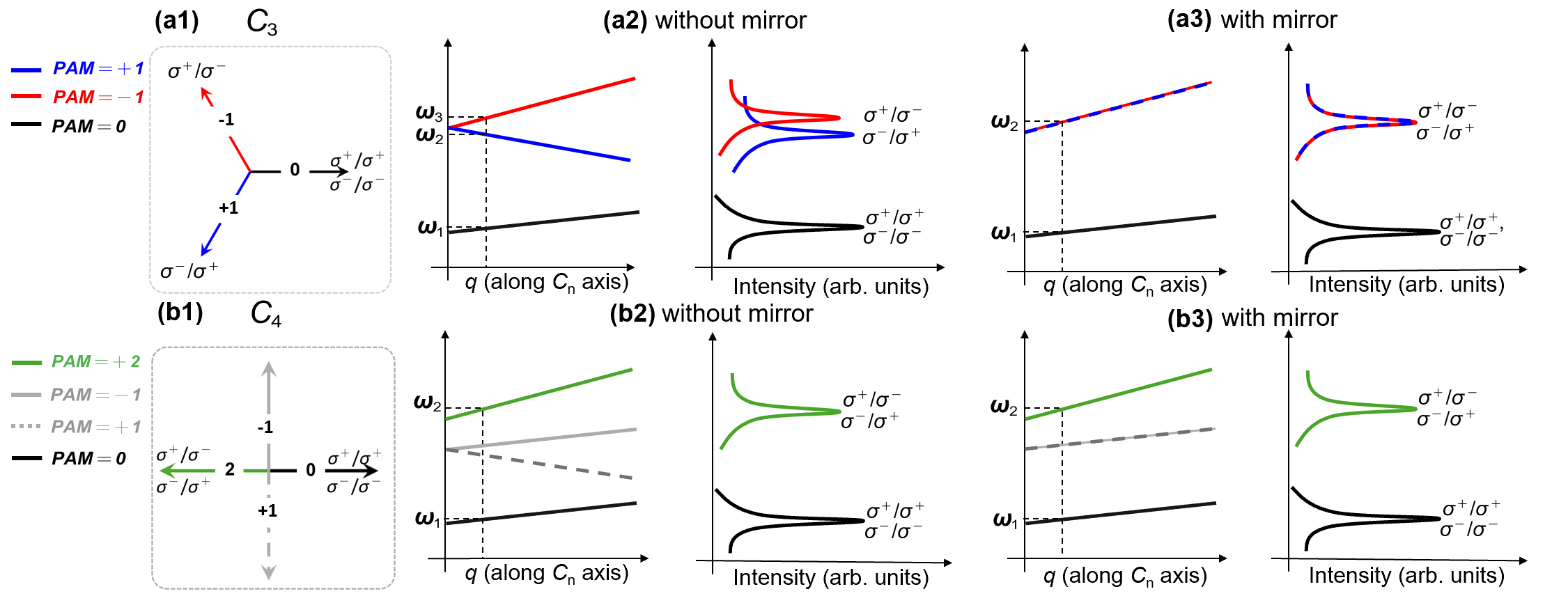}
  \caption{\label{fig:Raman_split_M} 
  Illustrations for the circularly polarized Raman scattering (Stokes process), along with the Raman shift in the systems with (a) $C_3$ and (b) $C_4$ rotation symmetries. 
  In the $C_3$-invariant systems, $\sigma^{+}/\sigma^{-}$ and $\sigma^{-}/\sigma^{+}$ Raman shifts split if there is no vertical mirror symmetry.
  Whereas for the $C_4$-invariant systems, $\sigma^{+}/\sigma^{-}$ and $\sigma^{-}/\sigma^{+}$ Raman shifts always degenerate regardless of the vertical mirror symmetry, since only the phonon mode with PAM=$2$ is active, exclusively in both $\sigma^+/\sigma^-$ and $\sigma^-/\sigma^+$ processes.
  }  
\end{figure*}

In CPRS at point $\boldsymbol {q}$, which belongs to the type-I magnetic group $C_3$, phonon modes with PAM = $\pm 1$ are nondegenerate.
In the Stokes process, the CPRS selection rule shown in Figure~\ref{fig:Raman_split_M} (a1) indicates that the phonon mode with PAM = $+1$ is active only in the $\sigma^-/\sigma^+$ process, while the phonon mode with PAM = $-1$ is active only in the $\sigma^+/\sigma^-$ process. Consequently, the corresponding Raman peaks split, as depicted in Figure~\ref{fig:Raman_split_M} (a2).

In contrast, the vertical mirror symmetry present in the little group of $C_{3v}$ enforces degeneracy between the phonon modes with PAM= $+1$ and $ -1$, as illustrated in Figure~\ref{fig:Raman_split_M} (a3).
The mirror-induced phonon degeneracy and the CPRS selection rule result in a degeneracy in the Raman peaks for both $\sigma^-/\sigma^+$ and $\sigma^-/\sigma^+$ processes. 
Same conclusion holds for $C_6$-invariant systems, where phonon modes with PAM = $\pm 2$ will be degenerate when the vertical mirror is present.

Note that in $C_{4}$-invariant systems, phonon modes with PAM=$+2$ are active in both the $\sigma^+/\sigma^-$ and $\sigma^-/\sigma^+$ processes (Figure~\ref{fig:Raman_split_M} (b1). Thus, the Raman splitting does not exist in the $C_4$ invariant system, regardless of the presence of the mirror symmetries, as shown in Figure~\ref{fig:Raman_split_M} (b2) and Figure~\ref{fig:Raman_split_M} (b3).

We also note that in $C_{n=4,6}$-invariant systems, phonon modes with PAM$=\pm 1$ are inactive in CPRS but active in infrared spectroscopy, serving as a complementary experimental technique to CPRS. 
The combination of CPRS and infrared spectroscopy provides an effective experimental approach to detect the presence of vertical mirror symmetry in these systems.

\subsection{$D_n$ point group}
\label{sec:Dn_group}

We now shift our attention to the type-I magnetic point group $D_n$, which encompasses systems characterized by multiple rotation axes, {i.e., a $n$-fold rotation symmetry $C_{n}$ and a twofold rotation symmetry $C_{2\perp}$ perpendicular to it.} We note that point groups $T$ and $O$, which similarly feature multiple rotation axes, will be addressed in Section~\ref{sec:TO_group}.

We assume the two perpendicular rotation axes in the point group $D_n$ to be $\boldsymbol {n}_1$ (for $C_n$) and $\boldsymbol {n}_2$ (for $C_{2\perp}$).
Based on the results in Section~\ref{sec:chiral_phonon_under_rotation}, AM must be parallel to both $\boldsymbol {n}_1$ and $\boldsymbol {n}_2$. {Due to $\boldsymbol {n}_1 \perp \boldsymbol {n}_2$,} the only feasible scenario is that phonon AM is zero. Thus, we conclude:

\textit{AM of the non-degenerated phonon is zero at $\boldsymbol {q}$ that belongs to the little point group of $D_n$. }

For the little type-I magnetic point groups of $D_{nd}$ and $D_{nh}$, which include additional mirror or inversion symmetries, they impose no further constraints on neither AM {nor additional degeneracies}. The relevant symmetry constraints remain within the framework of the type-I magnetic point groups $C_{nv}$ and $D_n$.

\subsection{$T$ and $O$ point groups }
\label{sec:TO_group}

The point group $T$ ($O$) consists of three mutually perpendicular $C_2$ ($C_4$) axes, along with four $C_3$ axes oriented along body diagonal directions. {These characterize the cubic symmetry.} 
In these two point groups, we define the PAM based on the eigenvalues of $C_{3[111]}$ along the body diagonal direction, as both point groups share this symmetry.
Based on the conclusions in Section~\ref{sec:Dn_group}, in the presence of multiple rotation axes that are not parallel to each other,
i.e., 

\textit{at $T$-/$O$-invariant momentum $\boldsymbol {q}$, the AM of the non-degenerated phonon is zero.}

\subsection{$T_{h}/T_d$ and $O_h$ point groups }

For $T_h$ and $O_h$ point groups, inversion symmetry ($\mathcal{P}$) is present but does not induce additional degeneracy beyond that of their rotation groups ($T$ and $O$), nor modify the AM. Consequently, all non-degenerate phonon modes in these groups exhibit zero AM.

\subsection{Time-reversal symmetry~($\mathcal{T}$)}
\label{sec:time_reversal_symmetry}

In previous sections, we explored symmetry constraints on chiral phonon-related properties under type-I magnetic point groups, where $\mathcal{T}$ is excluded, and the little point group of $\boldsymbol {q}$ contains no anti-unitary operations. This section focuses on symmetry constraints involving $\mathcal{T}$, particularly when $\boldsymbol {q}$ lies at time-reversal invariant momenta (TRIMs), corresponding to type-II magnetic little groups.

\subsubsection{Phonon chirality under $\mathcal{T}$}
\label{sec:trs_am_zero}

{In general, the time-reversal operator can be expressed as $\mathcal{T} = U\mathcal{K}$, where $\mathcal{K}$ denotes the complex conjugation operator and $U$ is a finite-dimensional unitary matrix, making $\mathcal{T}$ an anti-unitary operator.
For Bosons like phonon, $U$ should be the identity matrix, resulting in}
\begin{equation}
\begin{aligned}
\mathcal{T}=\mathcal{K},\\
\mathcal{T}^2=1. 
\end{aligned}
\end{equation}
{In this case, the anti-unitary operator has eigenvectors, associated with eigenvalues being arbitrary unitary complex numbers of $e^{i\phi}$~\cite{Anti_linear_Armin}.} 
Thus, under $\mathcal{T}$ symmetry, non-degenerate phonon modes at TRIMs exhibit zero AM and linear atomic motion. A detailed proof is shown in the Supporting Information Section S10.

\subsubsection{$\mathcal{T}$-extended little point groups}
\label{sec:little_group_with_T}

For type-II magnetic groups generated by type-I point group operations and $\mathcal{T}$, such as $C_{n}\&\mathcal{T}$, $C_{nh}\&\mathcal{T}, (n\neq 2)$, and others, the inclusion of $\mathcal{T}$ enforces the degeneracy of phonon modes with PAM $= \pm l_{ph}$. This arises from the theory of co-representation of magnetic groups, as these modes form a 2D IRREP of the corresponding magnetic group~\cite{el-batanouny_wooten_2008}.
For the point groups $T$/$T_{h}$, the $^1E_{(g/u)}$ and $^2E_{(g/u)}$ IRREPs combine to form a {2D} IRREP of the magnetic point group $T\mathcal{T}${/$T_{h}\mathcal{T}$}.
In these point groups, the AM of the non-degeneracy phonon is constrained to be zero by $\mathcal{T}$, while the AM of phonon modes belonging to high-dimensional IRREPs {can not be determined spontaneously, like the case in $C_{nv}$.} 
In the cases of {$C_{nv}$}, {$D_{n}$, $D_{nd}$, $D_{nh}$}, $T_d$, $O$ and $O_{h}$ point groups, $\mathcal{T}$ does not introduce additional degeneracy {or constraints on AM}.

\subsection{Phonon chirality under $\mathcal{PT}$ symmetry}
\label{sec:inversion_chiral}

This section explores the physical quantities and phenomena associated with chiral phonons under $\mathcal{PT}$ symmetry, given that $\mathcal{P}$ alone imposes no restrictions on AM since it is a pseudo-vector, as discussed previously. Under $\mathcal{PT}$ symmetry, two scenarios can be considered for $\boldsymbol {q}$: (1) When $\boldsymbol {q}$ is TRIM, it corresponds to type-II magnetic point groups; (2) When $\boldsymbol {q}$ is not located at TRIMs, it aligns with type-III magnetic point groups that break both $\mathcal{P}$ and $\mathcal{T}$ symmetries while preserving $\mathcal{PT}$ symmetry. In both scenarios, the {AM of the non-degenerate phonon mode} is zero~\cite{Sinisa_PT_AM_2023}.
A rigorous demonstration is in the Supporting Information Section S11,
which includes a more detailed discussion on the atomic motion belonging to different Wyckoff positions.

\subsubsection{PAM and CPRS under $C_n{\&}\mathcal{PT}$}
\label{sec:CPRS_under_PT}

Under $C_{n}{\&}\mathcal{PT}$ symmetry ($n\neq2$), phonon modes with PAM=$\pm l_{ph}$ are degenerate, forming a 2D IRREP of the corresponding magnetic group~\cite{el-batanouny_wooten_2008}. For non-degenerate modes, the AM is constrained to zero by $\mathcal{PT}$. 

The degeneracy of PAM$=\pm l_{ph}$ holds along the $C_n$-invariant $q$-path, highlighting a unique feature in the CPRS, where Raman shifts will also become degenerate, akin to the behavior observed in $C_{nv}$-invariant systems. 
However, in the $n = 4$ case, no Raman shifts splitting occurs, similar to the scenario discussed in Sec .~\ref {sec:cnv_3d}.
The observed degenerate or split patterns in the Raman shifts can offer valuable insights into the material's underlying symmetries.
As a result, CPRS combining infrared spectroscopy serves as a powerful tool for identifying the breaking of inversion symmetry in {$C_n$}-invariant systems. 

Consistent with earlier sections, we propose that external stimuli determine the AM of phonons excited in high-dimensional IRREPs during specific scattering processes. We illustrate this via a CPRS example using graphene's $G$ mode at $\boldsymbol {\Gamma}$ (see Supporting Information Section S12
).

\begin{figure*}
  \centering
  \includegraphics[width=1.0\textwidth]{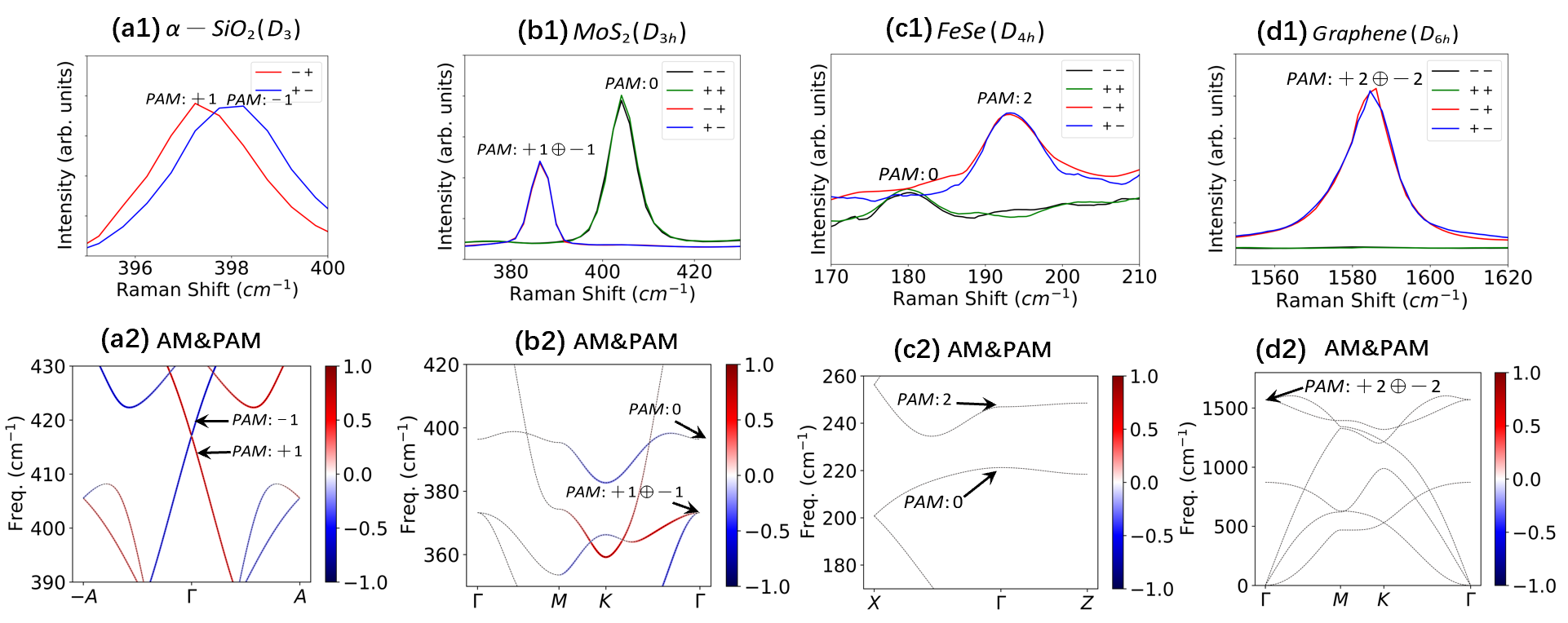}

  \caption{\label{fig:exp_raman}
  {
  CPRS experiments for materials with different symmetries. Since CRPS happens at $\boldsymbol {q}$ away from $\boldsymbol {\Gamma}$ point, the relevant symmetry could be lower than the point group of the crystal.}
  (a) Chiral crystal $\alpha-$SiO$_2$ with $D_3$ point group, {the relevant symmetry for CPRS at $\boldsymbol {q}$ is $C_3$}. (b) Monolayer-MoS$_2$ with an achiral structure and $C_{3v}$ symmetry {$D_{3h}$} point group, {the relevant symmetry for $\boldsymbol {q}$ is $C_{3v}$ with the additional $\mathcal{T}$ symmetry}. (c) FeSe with {$D_{4h}$} point group, {the relevant symmetry for $\boldsymbol {q}$ is $C_{4v}$ with the additional $\mathcal{PT}$ symmetry}. (d) Graphene with {$D_{6h}$}. {By dismissing the $z$-degree of freedom, the relevant symmetry for $\boldsymbol {q}$ is $C_{6v}$ with the additional $\mathcal{T}$ symmetry}. The first row shows the Raman shifts observed in CPRS, and the second row displays their corresponding phonon spectra, labeled with the phonon PAM and the $z$-component of AM. In (c1), the weak sample signal causes differing white-noise intensities between $\sigma^+/\sigma^-$ and $\sigma^-/\sigma^+$ configurations (and also $\sigma^{\pm}/\sigma^{\pm}$). We have globally scaled these intensities to equalize the background noise.
  BZ and the labels of the high-symmetry points are shown in the Supporting Information Figure S6.
  }
\end{figure*}

\subsection{Experimental verification by CPRS}
\label{sec:exp}
This section will focus on the experimental validation of our theoretical propositions by {CPRS}, combining the first-principle calculations.
Raman scattering occurs near $\boldsymbol {\Gamma}$ {point}, where the little group may differ from the crystal's point group, i.e., the point group at $\boldsymbol {\Gamma}$. We demonstrate diverse benchmarks in CPRS (Stokes process) using five materials with distinct symmetries, confirming the aforementioned theoretical results across different symmetry conditions.
Each material will be investigated under $\sigma^+/\sigma^+$, $\sigma^-/\sigma^-$, $\sigma^+/\sigma^-$, and $\sigma^-/\sigma^+$ four processes.

\subsubsection{CPRS at $\mathbf{q}$: $C_3$ vs. $C_{3v}$}
\label{sec:exp_C3C3v}
Section~\ref{sec:cnv_3d} established CPRS selection rules for degenerate phonon modes and their distinction between $C_3$ and $C_{3v}$. We demonstrate this using $\alpha$-SiO$_2$ ($C_3$-symmetric $\boldsymbol {q}$) and monolayer-MoS$2$ ($C_{3v}$-symmetric $\boldsymbol {q}$).

Figure~\ref{fig:exp_raman} (a1) reveals Raman shifts splitting for phonon modes with PAM$ =\pm1$ in $\alpha$-SiO$_2$, meanwhile, first-principles calculations confirm nonzero AM for these modes (Figure~\ref{fig:exp_raman} (a2)).
While in monolayer MoS$_2$, the Raman shifts for phonon modes with PAM = $\pm1$ are degenerate, with each mode active exclusively in the $\sigma^+/\sigma^-$ or $\sigma^-/\sigma^+$ process, as shown in Figure~\ref{fig:exp_raman} (b1). Both the CPRS data and numerical calculation match with our aforementioned theoretical results.
The atomic motions of the detected phonon modes are shown in the Supporting Information Section S13 and Figure S6.
Since all atoms in MoS$_2$ occupy $C_3$-invariant Wyckoff positions, phonon modes with PAM=$\pm$1 excited via CPRS exhibit AM=$\pm$1. This correspondence aligns with the symmetry analysis in Section~\ref{sec:AM_Cn_Wyckoff}.

\subsubsection{CPRS at $\mathbf{q}$ with $C_{4v}$\&$\mathcal{PT}$}

\label{sec:exp_C4}

Figure~\ref{fig:exp_raman} (c1) shows FeSe data ($D_{4h}$ at $\boldsymbol {\Gamma}$). At the CPRS detected $\boldsymbol {q}$, the little point group is $C_{4v}$ with $\mathcal{PT}$ symmetry.
Phonon modes with PAM $=2$ are Raman active in both the $\sigma^+/\sigma^-$ and $\sigma^-/\sigma^+$ processes, and phonon modes with PAM$ =0$ are Raman active in both the $\sigma^+/\sigma^+$ and $\sigma^-/\sigma^-$ processes, consistent with the theoretical results in Figure~\ref{fig:Raman_split_M} (b1). The phonon spectra in Figure~\ref{fig:exp_raman} (c2) show that the AM for all non-degenerate phonon modes is zero across the entire BZ, as constrained by the $\mathcal{PT}$ symmetry discussed in Section~\ref{sec:inversion_chiral}.

It should be noted that vertical mirror symmetry (or $\mathcal{PT}$) constrains the AM of PAM=2 phonon modes to zero, yielding nonzero-PAM/zero-AM modes that flip circular polarization. {This experimental result resolves two of the arguments presented in the introduction.} 

\subsubsection{CPRS at $\mathbf{q}$ with $C_{6v}$\&$\mathcal{PT}$}
\label{sec:exp_C6}

Figure~\ref{fig:exp_raman} (d1) shows graphene data ($D_{6h}$ at $\boldsymbol {\Gamma}$). Considering the 2D nature of graphene, neglecting the $z$-freedom reduces the point group to $C_{6v}$
Phonon modes with PAM$ =\pm2$ are Raman active in the $\sigma^-/\sigma^+$ and $\sigma^+/\sigma^-$ processes, in line with the theoretical analysis in Figure~\ref{fig:Raman_antiRaman}.
The phonon spectra for graphene shown in Figure~\ref{fig:exp_raman} (d2) indicate that the AM {of the nondegenerate phonons} is also zero across the entire BZ, due to the presence of $\mathcal{PT}$ symmetry in the first-principle calculation, aligning with the analysis in Section~\ref{sec:inversion_chiral}. The AM of the excited degenerate phonon is detailed in Supporting Information Section S12.
Since all atoms in graphene occupy $C_3$-symmetric Wyckoff positions, phonon modes with PAM=$\pm$2 excited via CPRS exhibit AM=$\mp$1. This correspondence aligns with the symmetry analysis in Section~\ref{sec:AM_Cn_Wyckoff}.

\begin{figure}
  \centering
  \includegraphics[width=0.5\textwidth]{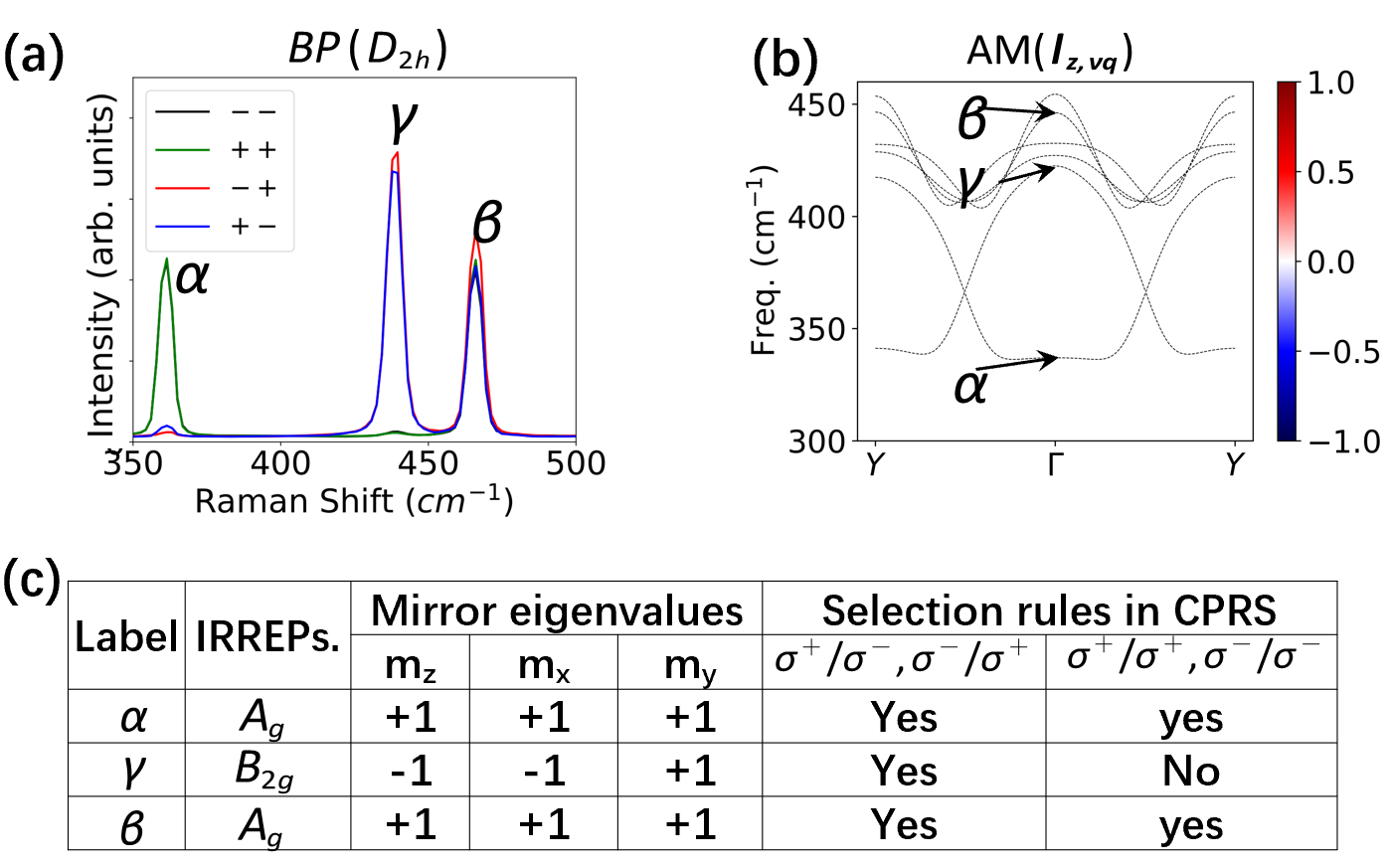}
  \caption{\label{fig:Exp_Mirror} 
  (a) CPRS results for the phonon modes $\alpha$, $\beta$, and $\gamma$ in black phosphorus. (b) The phonon spectra are labeled with phonon AM, demonstrating that the AM is zero for every phonon mode throughout the entire BZ. The phonon modes detected in the experiment are marked by arrows for clarity.
  (c) IRREPs, mirror eigenvalues selection rules for the phonon modes observed in the CPRS.
  }
\end{figure}

\subsubsection{Realization of the half-wave plate effect}

\label{sec:exp_Cs}

Section~\ref{sec:half_wave_plate} proposes the half-wave plate-like effect in systems with mirror symmetry (extendable to systems with multiple mirror planes). This requires phonon modes with odd mirror eigenvalues and CPL propagating parallel to the mirror plane.
To demonstrate this, we conducted a CPRS study on black phosphorus~(BP), which belongs to point group $D_{2h}$ with three mirrors perpendicular to each other, i.e., $\mathcal{M}_x$, $\mathcal{M}_y$ and $\mathcal{M}_z$. We set the CPL propagates along $y$-axis. For the $\boldsymbol {q}$ point where Raman scattering occurs, the corresponding point group is $C_{2v}$, characterized by two perpendicular mirror planes ($\mathcal{M}_{x}$ and $\mathcal{M}_z$) and a $C_2$ axis along the $y$-axis.

Raman tensor indicates that the $B_{2g}$ mode $\gamma$ ($\mathfrak{m}_z=\mathfrak{m}_x=-1$) will invert the circular polarization of the incident light when the Raman light propagates along the $y$-axis.
As shown in Figure~\ref{fig:Exp_Mirror} (a), the $B_{2g}$ mode $\gamma$ is Raman active only in the $\sigma^{+}/\sigma^{-}$ and $\sigma^{-}/\sigma^{+}$ processes, consistent with the selection rule shown in Figure~\ref{fig:Exp_Mirror} (c) and the theoretical conclusion in Section~\ref{sec:half_wave_plate}. 
Moreover, the AM of the Raman-excited phonon is zero due to the symmetry constraint by $C_{2v}$ (or $\mathcal{PT}$), as depicted in Figure~\ref{fig:Exp_Mirror} (b). The atomic motion of the detected phonon modes is detailed in the Supporting Information Figure S6.
This result confirms that a phonon mode with zero AM (i.e., an achiral phonon) can still flip the polarization of CPL.

\section{Conclusion and Discussion}
\label{sec:conclusion}

By systematically analyzing phonon AM, PAM, their relationship and associated physical phenomena across magnetic point groups, we comprehensively obtain the behavior of phonon chirality under different symmetries, establishing a foundation for future studies. Here are the key conclusions:

\begin{itemize}

    \item We demonstrate that AM and PAM generally lack an intrinsic relationship, but enumerate all specific conditions where a well-defined correspondence exists by symmetry analysis;

    \item We demonstrate that neither the nonzero PAM nor the cross-circular polarization scattering process is the indicator of chiral phonons;

    \item We discover novel half-wave plate-like phenomena, which is another way to flip the polarization of CPL by achiral phonons;
    
    \item We conduct experiments to validate our theory using five materials with distinct symmetries and benchmarks.

\end{itemize}

Supported by these theoretical and experimental results, our findings can help to identify the chirality of phonons through a range of experimental techniques.
Given the experimental challenges in directly probing AM, our theoretical analysis of phonon chirality under various symmetries, as well as associated experimental benchmarks, provides a practical framework for utilizing PAM to determine whether the corresponding AM vanishes. This approach facilitates the targeted search for materials hosting huge molecular Berry curvature or exhibiting substantial phonon magnetic moments, thermal Hall effect, etc.
Beyond CPRS, circularly polarized infrared/terahertz spectroscopy probes phonon PAM near $\boldsymbol{\Gamma}$ with absorption-based $C_n$-symmetric selection rules: phonon PAM must match incident photon PAM~\cite{chiral_ph_C4_2022, CoSnS_CP_2025}. Terahertz coherent excitation enables studies of Floquet states~\cite{Fan2024, Hubener2018, Kitamura2022, Chaudhary2020_PRR}, phonon magnetic moments~\cite{CeCl3_2022}, and field-driven chiral phonon manipulation~\cite{Baydin2022_PRB,zhang2024_CCDW}.

\section{Experimental and Computational  Methods}

Raman spectra were obtained using a Raman spectrometer (Horiba LabRAM HR Evolution) in a confocal backscattering configuration with a confocal pinhole of 200 $\mu$m. A 1 $\mu$m spot is obtained by focusing light from a 532 nm laser. The laser power does not exceed 150 $\mu$W and the integration time is 20 s.
The Raman spectra at room temperature are dispersed by 1800 gr/mm grating, and the backscattered signals are collected by a 50× objective lens. 
The materials under test were placed in an optical chamber with a high vacuum. 
The Raman spectral resolution is better than 1 cm$^{-1}$. The initial polarizer controls the polarization of the incident light. 
For the circular polarization configurations, the excited laser passes first through a vertical line polarizer and then through a half-wave plate placed in front of the objective lens, with rapid axial orientation at +45$^\circ$ and -45$^\circ$ to achieve
$\sigma^+$ and $\sigma^-$ circular incidence polarization's. Back-scattered Raman signals passing through the same half-wave plate are collected and analyzed employing a half-wave plate and a linear polarizer.

The phonon spectra calculation is implemented by VASP~\cite{kresse1996efficient, perdew_generalized_1996,kresse1999ultrasoft} and phonopy~\cite{phonopy} and the projector-augmented-wave (PAW) method~\cite{blochl1994projector,kresse1999ultrasoft} with the Perdew-Burke-Ernzerhof (PBE) exchange-correlation functional~\cite{perdew_generalized_1996} was used.
The plane-wave cutoff for kinetic energy was set as 500 eV for all materials.
The structures are fully relaxed and the force on each atom is less than 0.001 eV/\AA.
The supercell dimensions for $\alpha$-SiO$_2$, MoS$_2$, FeSe, graphene, and BP are set to $2\times2\times2$, $4\times4\times1$, $4\times4\times2$, $7\times7\times1$, and $2\times2\times3$, respectively. Correspondingly, the $k$-mesh configurations are $3\times3\times3$, $2\times2\times1$, $3\times3\times3$, $2\times2\times1$, and $2\times2\times2$, respectively.

\medskip
\section{Acknowledgements} \par 

We acknowledge the helpful discussion with Shuichi Murakami, Chen Fang, Hu Miao, Yuan Li, Bumjoon Kim, Yuanfeng Xu, and Yang Gao.
T. Zhang and S. Zhang acknowledge the support from National Key R\&D Project (Grant Nos. 2023YFA1407400 and 2024YFA1400036), and the National Natural Science Foundation of China (Grant Nos. 12374165 and 12447101).
T. Ying acknowledges the support from the National Natural Science Foundation of China (Grant No.52202342)

\section{Data Availability Statement} \par
The original data used in this work are available from the corresponding authors upon reasonable request.

%

\end{document}


\title{
Supplementary Materials of  ``Comprehensive Study of Phonon Chirality under Symmetry Constraints''
}

\author{Shuai Zhang}
\affiliation{Institute of Theoretical Physics, Chinese Academy of Sciences, Beijing 100190, China}
\author{Zhiheng Huang}
\author{Muchen Du}

\author{Tianping Ying}
\author{Luojun Du}
\affiliation{Institute of Physics, Chinese Academy of Sciences, Beijing 100190, China}
\author{Tiantian Zhang}
\email{ttzhang@itp.ac.cn} 
\affiliation{Institute of Theoretical Physics, Chinese Academy of Sciences, Beijing 100190, China}


\maketitle

\setcounter{section}{0}
\renewcommand{\thesection}{S\arabic{section}}

\onecolumngrid

\section{AM in the complex amplitude form}

\label{sec:supp_AM_AMP}

Now we prove that the AM of a phonon mode can be expressed in the form of the relative phase. The phonon wavefunction of the $\kappa$-th atom as:
\begin{equation}
    \boldsymbol{\epsilon}^\kappa_{\nu\boldsymbol{q}}=\{ A^{\kappa}_{x} e^{i\theta^{\kappa}_{x}}, A^{\kappa}_{y} e^{i\theta^\kappa_{y}}, A^{\kappa}_{z} e^{i\theta^\kappa_{z}}  \},
\end{equation}
where $A^\kappa_{\alpha}$ and $\theta^\kappa_{\alpha}$ are real numbers.
Thus, $l^\kappa_{\alpha, \nu\boldsymbol{q}}$ can be expressed as the relative phases between two degrees of freedom in \{$x,y,z$\}.
Let's take the $z$-component of $l_{\nu\boldsymbol{q}}^{\kappa}$ as an example:
\begin{equation}
\label{eq:relative_phase}
    \begin{aligned}
  l^\kappa_{z,\nu\boldsymbol{q}} &= \boldsymbol{\epsilon}_{\nu\boldsymbol{q}}^{\kappa\dagger} \cdot
               \begin{pmatrix}
                 0 & -i & 0\\
                 i & 0 & 0\\
                 0 & 0 & 0
               \end{pmatrix} \cdot 
               \boldsymbol{\epsilon}_{\nu\boldsymbol{q}}^{\kappa}  \\
            &= -A^\kappa_x A^\kappa_y e^{i (\theta^\kappa_y-\theta^\kappa_x)} i + A^\kappa_x A^\kappa_y e^{i(\theta^\kappa_x-\theta^\kappa_y)} i\\
            &= 2\text{Im}[A^\kappa_x A^\kappa_y e^{i (\theta^\kappa_x-\theta^\kappa_y)}].
    \end{aligned}
\end{equation} 
The other two $x,y$-components of AM are likewise and $l^\kappa_{\alpha,\nu\boldsymbol{q}}$ can be written in a general form:
\begin{equation}
    l^\kappa_{\alpha, \nu\boldsymbol{q}}= 2\text{Im}[A^{\kappa}_\beta A^{\kappa}_\gamma e^{i\epsilon_{\alpha(\beta\gamma)} (\theta^\kappa_\beta-\theta^\kappa_\gamma)}].
\end{equation}

\section{Extending discussion about helicity of a phonon}
\label{sec:supp_helicity_CCDW}

As we mentioned in the {main text}, the definition of chiral phonon based on AM depends on the reference direction.
Since $\boldsymbol{l}_{\nu\boldsymbol{q}}$ is a pseudo-vector, the helicity, an inner product of $\boldsymbol{q}$ and $\boldsymbol{l}_{\nu\boldsymbol{q}}$, is a pseudo-scaler and does not change under proper rotation operations. Thus, the definition for chiral phonons based on $h_{\nu\boldsymbol{q}}$ is convention-independent and well-defined. 

\begin{figure}[h]
  \centering
  \includegraphics[width=0.8\textwidth]{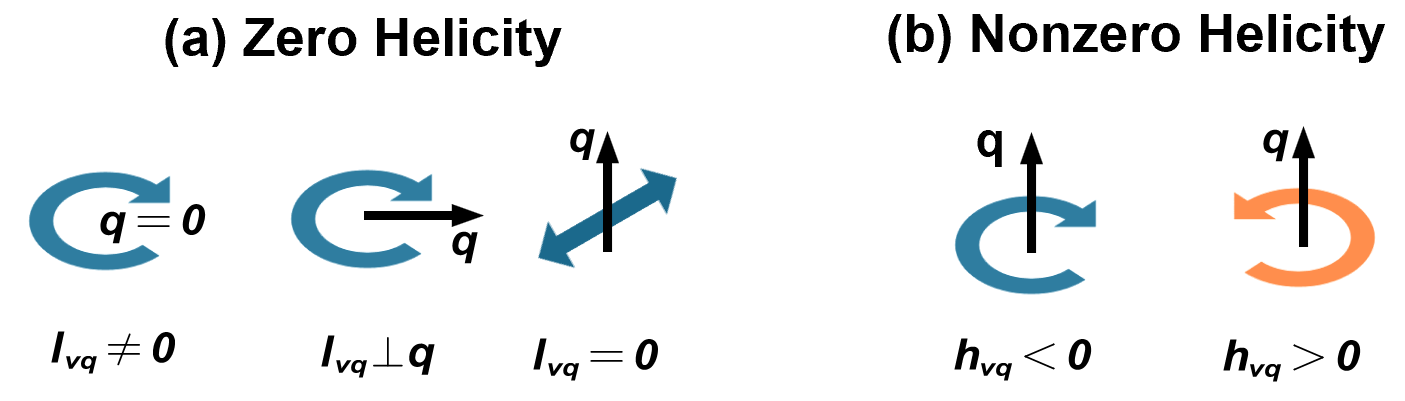}
  \caption{\label{fig:CP_helicity} 
    ``Chiral phonon'' is well-defined by phonon helicity, which is the inner product of the wavevector and AM of a phonon mode.  (a) Phonon modes with zero helicity have three possibilities, i.e.,  $\boldsymbol{q}$= 0, AM is perpendicular to $\boldsymbol{q}$, or AM is zero. (b) Phonon modes with nonzero helicity should have nonzero AM, and the sign of helicity shows the chirality of phonon modes. We note the AM and $\boldsymbol{q}$ do not have to be perfectly parallel.
}
\end{figure}

In the absence of symmetry constraints, there is no definitive relationship between AM and helicity, thus three scenarios arise for a phonon mode with zero helicity, as illustrated in Figure~\ref{fig:CP_helicity} (a): (i) $\boldsymbol{q} = 0$; (ii) $\boldsymbol{q} \neq 0$ and $\boldsymbol{l}_{\nu\boldsymbol{q}} \perp \boldsymbol{q}$; or (iii) $\boldsymbol{l}_{\nu\boldsymbol{q}} = 0$. In contrast, for a chiral phonon with nonzero helicity, it must possess nonzero AM, and the wave vector $\boldsymbol{q}$ cannot be perpendicular to the AM, as the representative case illustrated in Figure~\ref{fig:CP_helicity} (b), in which $\boldsymbol{q}$ is parallel to AM, and the sign of $h_{\nu\boldsymbol{q}}$ shows the chirality.

\begin{figure}[htp]
  \centering
  \includegraphics[width=0.5\textwidth]{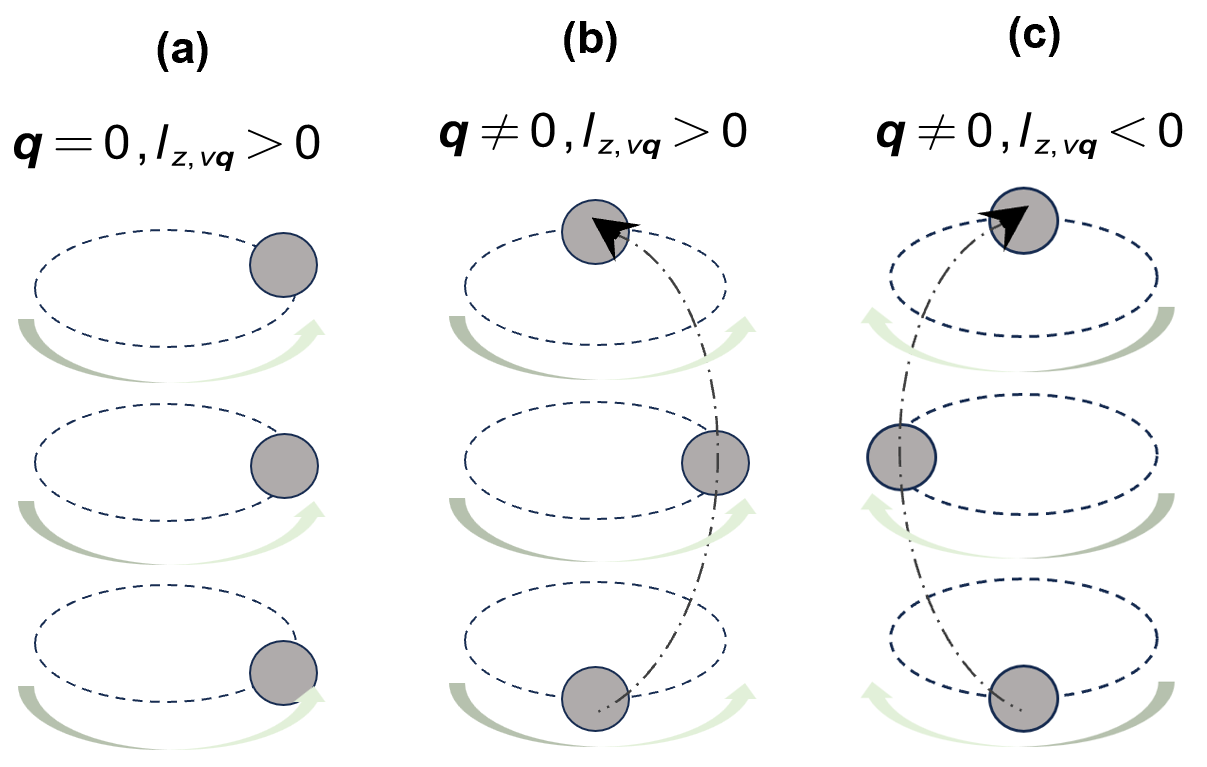}
  \caption{\label{fig:CP_CCDW} 
  A schematic illustration depicts the correlation between chiral phonons and chiral charge density waves (CCDWs). (a) A soft phonon mode at $\boldsymbol{\Gamma}$ with $l_{z,\nu\boldsymbol{q}}>0$ and zero helicity does not induce a screw structure phase transition, thereby failing to facilitate a CCDW phase. (b) and (c) present the two distinct chiral CCDW states induced by the soft chiral phonon modes, which possess $l_{z,\nu\boldsymbol{q}}>0$ and $l_{z,\nu\boldsymbol{q}}<0$, respectively, along with non-zero helicity, a consequence of the differing phases across various layers.
}
\end{figure}

Besides the convention independence, there is another benefit to defining the chiral phonon based on helicity, that is, it can be related to the concept of chiral CDW (CCDW), which represents the CDW phase has a screw structure. 
There are many underlying mechanisms proposed to induce the CCDW state, one of the most intriguing ones is based on the ``soft chiral phonon''.
Taking the CCDW phase along the $z$ direction as an example. The screw of the structure along the $z$ direction requires the rotation of the atoms in each of the crystal plane perpendicular to $z$ (i.e., the $l_{z, \nu\boldsymbol{q}}$ is not zero) and there should be a phase different between different layers ($\boldsymbol{q}_z$ should not be zero), as shown in Figure~\ref{fig:CP_CCDW}.
 
\section{The PAM conserving scattering process which only involves the phonon emitting/absorption process.}
\label{sec:supp_q_path_gamma}

Phonon emission/absorption processes exclusively change phonon number ($\pm$1) without creating other excitations (e.g., electrons, holes, or excitons), such as Raman scattering or infrared absorption.
If PAM conservation is satisfied in these scattering processes, both the incident/scattered light and the final excited states should be the eigenstates of $C_n$.
Note that if the incident/scattered light is the eigenstate of $C_n$, the propagating direction of them should parallel to the rotation axis.
Now we suppose a phonon mode $\epsilon_{\nu\boldsymbol{q}}$ is a eigenstate of $C_n$, $\boldsymbol{q}$ should satisfy $C_n \boldsymbol{q}=\boldsymbol{q+G}$,  $\boldsymbol{G}=\sum_{i}n_i \boldsymbol{b}_i$, where $\boldsymbol{b}_i$ are the reciprocal lattice vectors, and $n_i\in \mathbb{Z}$.
Note that if $n_i\neq0$, the $\boldsymbol{q}$ is on a $C_n$ axis which does not path through the $\boldsymbol{\Gamma}$ point (such as the $\boldsymbol{K}$ point of graphene).
If the scattering process involves such a phonon, the propagating direction of the scattered light is not along the $C_n$ axis, and it is impossible to be an eigenstate of $C_n$. In this case, the PAM conservation is meaningless.
\begin{figure}
  \centering
  \includegraphics[width=0.90\textwidth]{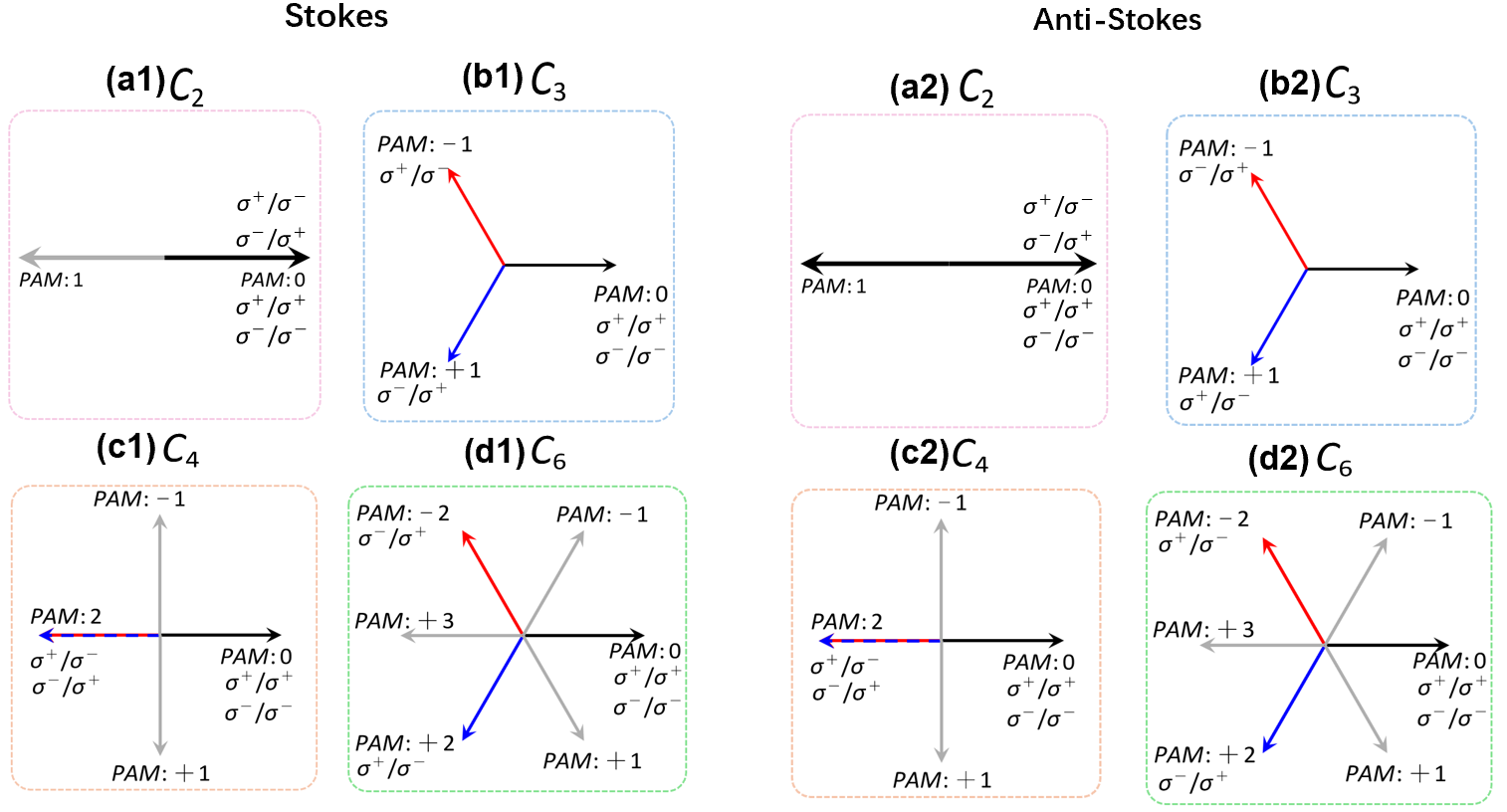}
  \caption{\label{fig:s_rule}
The selection rules for circularly polarized Raman scattering in the Stokes (a1-d1) and anti-Stokes (a2-d2) processes for the systems with $C_n={2,3,4,6}$ rotation symmetries, where the incident/scattered lights propagate along the rotation axis. 
Each of the arrows represents the eigenvalue of a phonon mode in the complex plane, associated with its PAM value. 
} 
\end{figure}

\section{CPRS election rules for the Stokes/anti-Stokes process}
\label{sec:supp_CPRS_selec}

Firstly, we consider the Stokes process, which corresponds to the scattering process that emits a phonon. As mentioned in the main text, we set the propagating direction of light parallel to the rotation axis.
In this case, CPLs with $\sigma^+=(1,i,0)$ and $\sigma^-=(1, -i, 0)$ are the eigenvectors of $C_{n=3,4,6}$, the corresponding PAM of the is $+1$ and $-1$ respectively.
By dismissing the inter-media virtual states, the initial state of the Stokes process corresponds to the incident light, i.e. $|{I}\rangle=|{l_i}\rangle$, and the final state is the direct product of the scattered light and the emitted phonon, i.e., $|S\rangle=|l_s\rangle\otimes |l_{ph}\rangle$. Here, $l_{s}$, $l_{i}$, and $l_{ph}$ represent the PAM of the scattered light, incident light, and the emitted phonon, respectively.
The initial state and the final state should have the same eigenvalues for $C_n$, i.e.:
 \begin{equation}
    \label{eq:Cn_conserve}
    e^{\frac{-i 2 \pi l_i}{n}}  = e^{\frac{-i 2 \pi l_s}{n}} \cdot e^{\frac{-i 2 \pi l_{ph}}{n}},
\end{equation}
Thus, in addition to the conservation rules for energy and momentum, PAM contributes an {additional} selection rule in the CPRS (Stokes process):
\begin{equation}
\label{eq:C3_pam_conserve}
    l_{i} - l_{s} = l_{ph} \quad \text{modulo} \quad n.
\end{equation}

Figure~\ref{fig:s_rule} (a1)-(d1) show the selection rules for the CPRS in the Stokes process. Each arrow represents the eigenvalue of a phonon mode on the complex plane, associated with its PAM value. 
For the anti-Stokes process, which absorbs a phonon, the initial state is $|{I}\rangle=|{l_i}\rangle\otimes|l_{ph}\rangle$, and the final state is $|S\rangle=|l_s\rangle$.
Based on the previous discussion, the CPRS selection rule for the anti-Raman process is shown in Fig~\ref{fig:s_rule} (a2)-(d2).
For the $C_2$ case, in the CPRS process, $\sigma^+$ and $\sigma^-$ are has the $-1$ eigenvalue of $C_2$, i.e., the PAM=$+1$ (or $-1$, equivalently after mod 2.). Thus, for all four CPRS processes, only the phonon mode with PAM=0 is active.

In systems with non-symmorphic rotation symmetries, where PAM is not necessarily an integer, the CPRS selection rule is based on the projective representation of $C_{n, \boldsymbol{\tau}_{m/n}}$, i.e., 
\begin{equation}
    \begin{aligned}
    h[\mathcal{D}(C_{n, \boldsymbol{\tau}_{m/n}}) ]&= \mathcal{D}(C_{n, \boldsymbol{\tau}_{m/n}}) / e^{-i \boldsymbol{q}\boldsymbol{\tau_{m/n}}}\\
    &=\mathcal{D}(C_n),
    \end{aligned} 
\end{equation}
and the projective PAM, i.e.,  $l_{rot}$ is defined by
\begin{equation}
    h[D(C_{n, \boldsymbol{\tau}_{m/n}}) ] \epsilon_{\nu\boldsymbol{q}}=e^{-i \frac{2 \pi l_{rot}}{n}} \epsilon_{\nu\boldsymbol{q}}.
\end{equation}
corresponds to the pure rotation component of PAM. 

\section{CPRS vs Raman tensor}

\label{sec:supp_Stokes_antiStokes}
In the main text, we have mentioned that the CPRS can distinguish the Raman/anti-Raman process. Here we give a detailed discussion about it. We suppose the light propagates along the $C_{6z}$ axis.
The Raman tensor of the phonon mode with PAM = $-2$ is
\begin{equation}
    R(l_{ph}=-2) =
    \begin{pmatrix}
        e & f  &0\\
        f & -e &0\\
        0 & 0  &0
    \end{pmatrix},
\end{equation}
and the intensity of the $\sigma^-/\sigma^+$ and $\sigma^+/\sigma^-$ process reads
\begin{equation}
    \begin{aligned}
        I(\sigma^+/\sigma^-)=|2(e-if)|^2,\\
        I(\sigma^-/\sigma^+)=|2(e+if)|^2.
    \end{aligned}
\end{equation}
Namely, both the $\sigma^-/\sigma^+$ and $\sigma^+/\sigma^-$ processes are active in the CPRS. However, for the Stokes process in Figure 2(b) (in the main text), which has a higher intensity, the phonon mode with PAM = $-2$ is active only in the $\sigma^-/\sigma^+$ process. This distinction is crucial for explaining the Raman splitting observed in experiments. Consequently, the Raman tensor alone cannot differentiate between the two processes or identify the active one.

\section{The tight-binding model illustrating that the phonon mode with the same PAM can have opposite AM}
\label{sec:supp_TB_C61}

\begin{figure}
  \centering
  \includegraphics[width=0.9\textwidth]{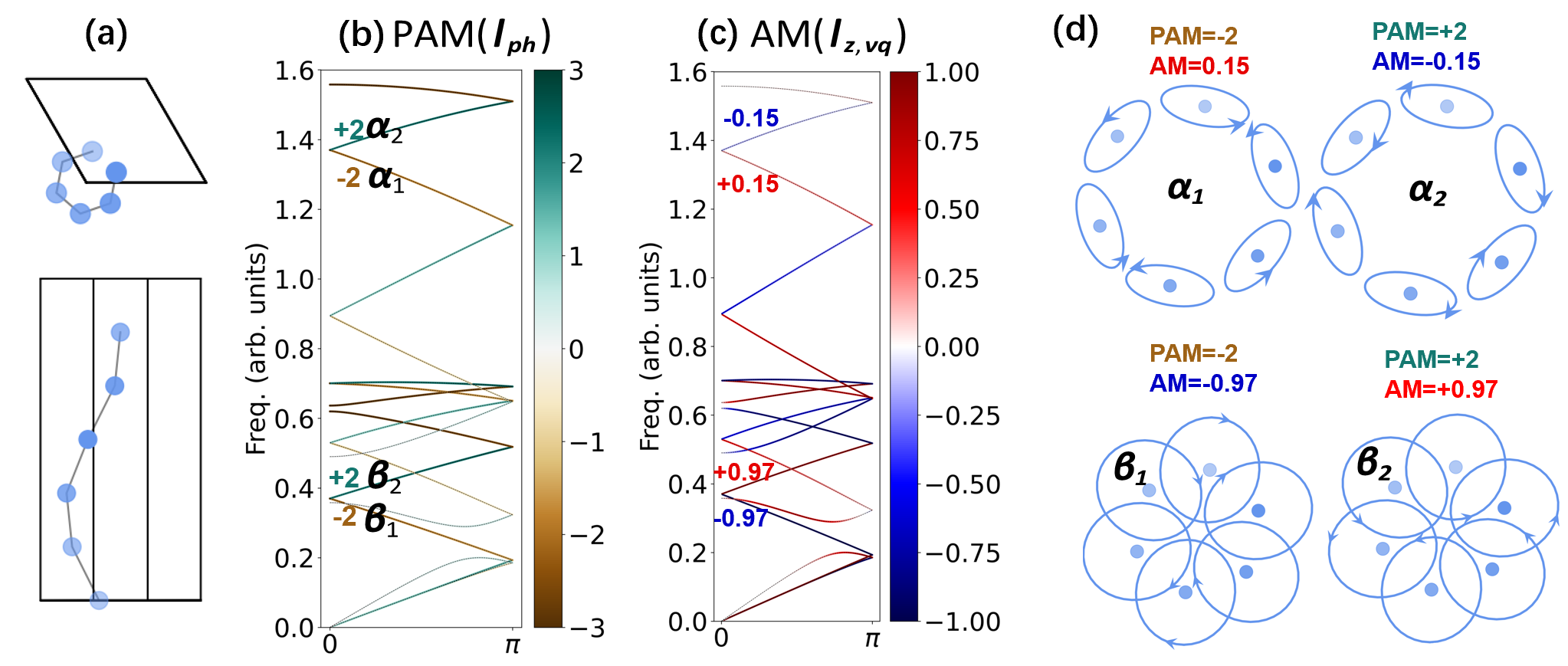}
  \caption{\label{fig:C61_pam_am}
  {AM, PAM and atomic motions in systems with $C_{6,\boldsymbol{\tau}_{1/6}}$ screw rotation symmetry.}
(a) Top and side views for the lattice with $C_{6,\boldsymbol{\tau}_{1/6}}$ screw symmetry.
(b) PAM and (c) $z$ component AM ($l_{z,\nu\boldsymbol{q}}$) of phonon modes along $k_z$ direction.
(d) Atomic motions in the $x-y$ plane for the phonon modes $\alpha_{1,2}$ and $\beta_{1,2}$.}
\end{figure}

Next, we illustrate that phonon modes with nonzero PAM can also exhibit zero AM using a tight-binding (TB) model, describing a quasi-one-dimensional chain with $C_{6,\boldsymbol{\tau}_{1/6}}$ screw rotation symmetry ($P6_1$, No. 169 SG group). The crystal structure of the system is depicted in Figure~\ref{fig:C61_pam_am} (a),
{in which the lattice constant are $\boldsymbol{a}=\boldsymbol{b}=2$, and $\boldsymbol{c}=6$. The atom positions (in basis of the lattice vectors) are shown in Table~\ref{tab:C6_1}. Here, we only considered the nearest ``spring'' constant,  with the potential energy in the form of $V_{\boldsymbol{\tau}}=V_l(\boldsymbol{\tau})+V_t(\boldsymbol{\tau})$.
Here, $\boldsymbol{\tau}$ represents the vector connected by a spring.
$V_l(\boldsymbol{\tau})$ and $V_t(\boldsymbol{\tau})$ are the longitude and the transverse term with the form:}
\begin{align}
  V_l(\tau)=\frac{1}{2}*L*|\boldsymbol{\tau}\cdot(\boldsymbol{u_2}-\boldsymbol{u_1}) |^2 \\
  V_t(\tau)=\frac{1}{2}*T*|\boldsymbol{\tau}\times(\boldsymbol{u_2}-\boldsymbol{u_1}) |^2,
\end{align}
{where $\boldsymbol{u}_1$ and $\boldsymbol{u_2}$ is the displacement away from the equilibrium positions. The longitudinal spring constant $L$ is set to be 0.5, and the transverse spring constant $T$ is set to be 0.1, respectively.}

The phonon spectra labeled with PAM and AM are shown in Figure~\ref{fig:C61_pam_am} (b) and (c), respectively.
Phonon modes $\alpha_1$ and $\beta_1$, located at $\boldsymbol{q}=(0, 0, 10^{-4}\pi)$, exhibit identical PAM ($=-2$) but opposite AM ($+0.15$ for $\alpha$ and $-0.97$ for $\beta$). Their corresponding atomic motions are shown in Figure~\ref{fig:C61_pam_am} (d), indicating that the phonon mode with larger AM has the atomic motion with larger circular polarization.
If the force constant is modulated without changing the symmetry of the system, phonon modes $\alpha_1$ and $\beta_1$ can hybridize and then form new phonon modes with nonzero PAM but zero AM (PAM $=-2$, AM $=0$ ), since they belong to the same irreducible representation (IRREP).

\section{The representation matrix of $C_n$ for the phonon system}
\label{sec:supp_PAM_orbital_spin}

The PAM, including its spin and orbital components, arises from the representation of the $C_n$ symmetry. In the general case, the representation matrix for $C_n$ {at $\boldsymbol{q}$} can be obtained by applying it to the basis vector $\boldsymbol{\epsilon}_{\kappa\alpha, \boldsymbol{q}}$, which corresponds to the Bloch sum of the atomic displacements.
$\epsilon_{\kappa\alpha, \boldsymbol{q}}=\sum_{l} e^{i\boldsymbol{q}\cdot(\boldsymbol{R}_{l}+\boldsymbol{\tau}_{\kappa})} \epsilon_{\kappa\alpha}(\boldsymbol{R}_l+\boldsymbol{\tau}_{\kappa})$, and the corresponding transformed state is:

\begin{equation}
    \begin{aligned}
        C_{n} {\epsilon}_{\kappa\alpha, \boldsymbol{q}}  = &C_{n} \sum_{l} e^{i\boldsymbol{q}\cdot(\boldsymbol{R}_{l}+\boldsymbol{\tau}_{\kappa})} \boldsymbol{\epsilon}_{\kappa\alpha}(\boldsymbol{R}_l+\boldsymbol{\tau}_{\kappa}) 
    \end{aligned}
\end{equation}
After setting $C_n (\boldsymbol{R_l}+\boldsymbol{\tau_{\kappa}})=\boldsymbol{R}_{l'}+\boldsymbol{\tau_{\kappa'}}$, we have:

\begin{equation}
    \begin{aligned}
        C_{n} {\epsilon}_{\kappa\alpha, \boldsymbol{q}}  
         & =   \sum_{l'}  e^{i C_{n} \boldsymbol{q} \cdot (\boldsymbol{R}_{l'} +\boldsymbol{\tau}_{\kappa'})} \sum_{\beta}C_{n, \alpha\beta} 
        \epsilon_{\kappa'\alpha}(\boldsymbol{R}_{l'} + \boldsymbol{\tau}_{\kappa'}) \\
        & =   \sum_{l, \beta}  e^{i C_{n} \boldsymbol{q} \cdot (\boldsymbol{R}_{l} +\boldsymbol{\tau}_{\kappa'})}  C_{n, \alpha\beta} 
        \epsilon_{\kappa'\alpha}(\boldsymbol{R}_{l} + \boldsymbol{\tau}_{\kappa'}) 
    \end{aligned}
\end{equation}
Here, we substitute $l'\longrightarrow l$ for the periodic-boundary condition, and $C_n$ is the representation matrix of rotation in the Euclidean space. Since $\boldsymbol{q}$ is $C_n$-invariant, thus we have:
\begin{equation}
    \begin{aligned}
        C_{n} \boldsymbol{q}=\boldsymbol{q} + \boldsymbol{G}. 
    \end{aligned}
\end{equation}
The reciprocal lattice vector, denoted as $\boldsymbol{G}$, is defined as the sum of the integer multiples of the basis vectors of the reciprocal lattice, $\boldsymbol{G} = \sum_{i} n_i \boldsymbol{b}_i$, where the coefficients $n_i$ vary depending on the specific case. Following the operation by $C_n$, the resulting state can be represented as:
\begin{equation}
    \begin{aligned}
        C_{n} \epsilon_{\kappa\alpha, \boldsymbol{q}} 
        & = e^{i \boldsymbol{G}\cdot\boldsymbol{\tau}_{\kappa'}} P_{\kappa'\kappa} \sum_{\beta}C_{n, \alpha\beta} \epsilon_{\kappa'\alpha,\boldsymbol{q}}, 
    \end{aligned}
\end{equation}
where $P_{\kappa'\kappa}$ denotes a permutation matrix that describes the transformation of the $\kappa$-th atom to the $\kappa'$-th atom within the primitive cell. Upon left-multiplying $u_{{\kappa' \beta},\boldsymbol{q}}$ by this matrix, the representation matrix for the symmetry operation $C_{n}$ is obtained as follows:
\begin{equation}
\label{eq:supp_Cn_repmat}
    D(C_{n})_{\kappa'\beta, \kappa\alpha}=e^{i \boldsymbol{G}\cdot\boldsymbol{\tau}_{\kappa'}} P_{\kappa'\kappa} C_{n, \alpha\beta}.
\end{equation}
Then, PAM can be expressed as $l_{ph}$ through the eigenvalue of, i.e., 
\begin{align}
       \mathcal{D}(C_n) u_{\nu\boldsymbol{q}}  = e^{-i 2 \pi l_{ph}/n} u_{\nu\boldsymbol{q}}
\end{align}
where $\nu$ is the index of a phonon mode.

There might be a misunderstanding that the PAM can always be decoupled into the spin and orbital parts. However, orbital and spin PAM are not always well-defined. Based on the representation of $C_n$, we will give a detailed discussion below.

Now we suppose that there are two sublattices in the primitive cell, and occupy the $C_n$ invariant Wyckoff positions.
In this case, $D(C_n)$ has the form of
\begin{equation}
\label{eq:supp_C3_rep}
    D(C_n)= 
    \begin{pmatrix}
        e^{i \bm{G} \cdot \bm{\tau}_1} \cdot C_{n} & 0 \\
        0 & e^{i \bm{G} \cdot \bm{\tau}_2} \cdot C_{n},
    \end{pmatrix}
\end{equation}
In Eq.~\ref{eq:supp_C3_rep},
The matrix $D(C_n)$ is block diagonal and can be block diagonalized separately with respect to the 1st and 2nd sublattices, i.e., the eigenvalues/eigenvectors of $e^{i \bm{G}_1 \cdot \bm{\tau}_1} \cdot C_{n}$ or $e^{i \bm{G} \cdot \bm{\tau}_2} \cdot C_{3}$ is also the eigenvalues/eigenvectors of $D(C_n)$.
In this case, the phase factors $e^{i \bm{G} \cdot \bm{\tau}_1}$ and $e^{i \bm{G}_1 \cdot \bm{\tau}_2}$ contribute the ``orbital part'' of the PAM, and the eigenvalue of $C_{3}$ contribute the ``spin'' part of the PAM.

However, if these two sublattices do not occupy the $C_n$ invariant Wyckoff positions, the $D(C_n)$ is not block diagonalized, and has the form of
\begin{equation}
    D(C_n)= 
    \begin{pmatrix}
       0 & e^{i \bm{G} \cdot \bm{\tau}_1} \cdot C_{n}  \\
      e^{i \bm{G} \cdot \bm{\tau}_2} \cdot C_{n} & 0
    \end{pmatrix}
\end{equation}
In this case, $D(C_n)$ can not be diagonalized for different sublattices separately, thus there is also no appropriate method to define the orbital/spin part of the PAM, and attempting to decouple these components is rendered devoid of meaning.

\section{Proof of ``If there is only one mirror plane, AM is perpendicular to it''.}
\label{sec:supp_AM_under_mirror}

In systems exhibiting only one mirror plane parallel to the $z$ axis ($\mathcal{M}_{\parallel z}$), atoms can be categorized into two distinct classes. In class I, atoms $a_\kappa$, $b_\kappa$ are related by $\mathcal{M}_{\parallel z}$; while in class II, atoms $o$ are located at $\mathcal{M}_{\parallel z}$-invariant {Wyckoff positions in real space}. Phonon modes at $\mathcal{M}_{\parallel z}$-invariant momenta can be expressed as
\begin{equation}
    \begin{aligned}
      \boldsymbol{\epsilon}_{\nu\boldsymbol{q}} &=\{ \boldsymbol{\epsilon}^{a_\kappa}, \boldsymbol{\epsilon}^{b_\kappa}, \boldsymbol{\epsilon}^o\}, 
    \end{aligned}
\end{equation}
with mirror eigenvalue of $\mathfrak{m}_{\parallel z} = e^{i\phi}$ ($\phi=0$ or $\pi$).
We note that $\mathcal{M}_{\parallel z}$ cannot mix components from different classes. For a general wavefunction, each class must be an eigenvector of $\mathcal{M}_{\parallel z}$, possessing the same $\mathfrak{m}_{\parallel z}$. Without loss of generality, we assume $\mathcal{M}_{\parallel z}$ to be $\mathcal{M}_{x}$, then we have
\begin{equation}
\label{eq:ph_mirror_opt}
    \begin{aligned}
      \mathcal{M}_{x} \boldsymbol{\epsilon}_{\nu\boldsymbol{q}} &=
      e^{i\phi}\{ \boldsymbol{\epsilon}^{a_\kappa}, \boldsymbol{\epsilon}^{b_\kappa}, \boldsymbol{\epsilon}^o\} \\
      &=
     e^{i\phi} \{\epsilon_x^{a_\kappa}, \epsilon_y^{a_\kappa}, 
     \epsilon_z^{a_\kappa},
     \epsilon_x^{b_\kappa}, \epsilon_y^{b_\kappa}, 
     \epsilon_z^{b_\kappa}, 
     \epsilon_x^o, 
     \epsilon_y^o,
     \epsilon_z^o\}\\
     &=
     \{-\epsilon_x^{b_\kappa}, \epsilon_y^{b_\kappa},
     \epsilon_z^{b_\kappa},
     - \epsilon_x^{a_\kappa}, \epsilon_y^{a_\kappa},
     \epsilon_z^{a_\kappa}, 
     -\epsilon_x^o, 
     \epsilon_y^o,
     \epsilon_z^o
     \}.
    \end{aligned}
\end{equation}
Based on the equations in Eq.~\ref{eq:ph_mirror_opt},
the amplitude of $a_\kappa$ and $b_\kappa$ sublattices should be same, i.e., $A^{a_\kappa}_\alpha=A^{b_\kappa}_\alpha$,
{and} the relationship of $\theta_\alpha$ for the $a_\kappa$ and $b_\kappa$ sublattices should be: 
\begin{equation}
    \begin{aligned}
        \theta_{x}^{b_\kappa}&=\theta_{x}^{a_\kappa} + \phi + (2n+1)\pi\\ 
        \theta_{y}^{b_\kappa}&=\theta_{y}^{a_\kappa} + \phi.
    \end{aligned}
\end{equation}
For systems belong to class I, the $z$-component of AM can be expressed as
\begin{equation}
    \begin{aligned}
        l^{a_\kappa}_{z, \nu\boldsymbol{q}}&=2\text{Im} [A^{a_\kappa}_xA^{a_\kappa}_ye^{i(\theta^{a_\kappa}_x-\theta^{a_\kappa}_y)}], \\
        l^{b_\kappa}_{z,\nu\boldsymbol{q}}&=2\text{Im} [A^{a_\kappa}_xA^{a_\kappa}_ye^{i(\theta^{a_\kappa}_x-\theta^{a_\kappa}_y+(2n+1)\pi)}],
    \end{aligned}
\end{equation}
and hence $l^{a_\kappa}_{z}=-l^{b_\kappa}_{z}$, i.e., the total $z$-component of AM is zero in class I. 

For systems belong to class II, we have
\begin{equation}
    \begin{aligned}
    \phi + \theta^o_x &= (2n+1)\pi + \theta^o_x, \\ 
    \phi + \theta^o_y &= \theta^o_y, 
    \end{aligned}
\end{equation}
where $n$ is an integer. If $\phi=\pi$, then $\theta_y$ has no root, Eq.~\ref{eq:ph_mirror_opt} holds only when $A^o_y=A^o_y=0$; if $\phi=0$, then $\theta_x$ has no root, Eq.~\ref{eq:ph_mirror_opt} holds only when $A^o_x=A^o_x=0$. Thus, we get $l^o_{z, \nu\boldsymbol{q}}=0$, and the total $z$-component of AM $l_{z,\nu\boldsymbol{q}}=l^{a_\kappa}_{z,\nu\boldsymbol{q}}+l^{b_\kappa}_{z,\nu\boldsymbol{q}}+l^{o}_{z,\nu\boldsymbol{q}}$ should also be zero.
Similarly, the $y$-component also vanishes. These demonstrate that the AM is oriented perpendicular to the mirror plane {at the $\mathcal{M}$-invariant momenta}.

\section{An intuitive picture about the ``half-wave plate-analogous effect''}
\label{sec:supp_half_wave_plate}

For the $B$ mode ($\mathfrak{m}=-1$) of phonon under $\mathcal{M}_z$, as shown in Figure 3 of the main text, the Raman tensor of it reads:
\begin{equation}
    R(B) =
    \begin{pmatrix}
        0 & 0  &e\\
        0 & 0  &f\\
        e & f  &0
    \end{pmatrix},
\end{equation}
{We denote the right(left)-handed circularly polarized light (R-CPL/L-CPL), $|{+}\rangle = (1, 0, i)^T$ ($|{-}\rangle = (1, 0, -i)^T$), propagating along the $y$-direction. The scattering intensity for the $\sigma^+/\sigma^-$ process reads:}
\begin{equation}
\begin{aligned}
    I(+/-) &= |\langle{+}|R(B)|{-}\rangle|^2 \\
    & = |\begin{pmatrix}
        1, 0, -i
    \end{pmatrix} 
    \begin{pmatrix}
        0 & 0  &e\\
        0 & 0  &f\\
        e & f  &0
    \end{pmatrix}
    \begin{pmatrix}
        1\\
        0\\
        -i
    \end{pmatrix}|^2\\
    &=|-2ei|^2,
\end{aligned}
\end{equation}
{and the $\sigma^+/\sigma^+$ process reads:}
\begin{equation}
\begin{aligned}
    I(+/+) &= |\langle{+}|R(B)|{-}\rangle|^2 \\
    & = |\begin{pmatrix}
        1, 0, -i
    \end{pmatrix} 
    \begin{pmatrix}
        0 & 0  &e\\
        0 & 0  &f\\
        e & f  &0
    \end{pmatrix}
    \begin{pmatrix}
        1\\
        0\\
        i
    \end{pmatrix}|^2\\
    &=0.
\end{aligned}
\end{equation}
{Similarly, the intensity for the $I(-/+)$ and $I(-/-)$ are $|2ei|^2$ and $0$. These results indicate that only the cross-polarization is promised for $B$ modes.}

{To give an intuitive picture of this phenomenon, we consider this process under the conservation of the mirror eigenvalues under the scattering process.}
Consider an incident R-CPL, $|{l_i}\rangle = (1, 0, i)^T$, propagating along the $y$-direction (parallel to the $\mathcal{M}_z$ plane). Under mirror symmetry $\mathcal{M}_z$, the R-CPL can be decomposed into two linearly polarized components: $|{x}\rangle=(1, 0, 0)^T$ with $\mathfrak{m}_z=+1$ and $|{z}\rangle=(0, 0, i)^T$ with $\mathfrak{m}_z=-1$.
Next, we analyze the scattering processes for each component. For the $|{x}\rangle$ component, the scattered light must share the mirror eigenvalue of the combined state $|{x}\rangle\otimes|{B}\rangle$, yielding $\mathfrak{m}_z$ = $+1\times-1=-1$, consistent with $\mathfrak{m}_z$ conservation. As a result, the initial $|{x}\rangle$ component transforms into $z$-linearly polarized light, $|{z'}\rangle=(0,0,1)^T$.
Similarly, the initial $|{z}\rangle$ component transforms into $x$-linearly polarized light, yielding $|{x'}\rangle=(i, 0, 0)^T$. Notably, the $y$ component, with $\mathfrak{m}_z=+1$, is forbidden by quantum gauge field theory, as free light can only have perpendicular polarization.
Thus, the incident R-CPL, $|{l_i}\rangle=(1, 0, i)^T$, is scattered into L-CPL, $|{l_s}\rangle=(1, 0, -i)^T$.
We emphasize that this analysis does not constitute a rigorous proof. Throughout our derivation, we assumed incident and scattered photons share identical initial phases, a condition imposed to ensure consistency with the Raman tensor formalism. For instance, the transition $|x\rangle = (1,0,0)^T \rightarrow |z'\rangle = (0,0,1)^T$ was modeled without phase accumulation. In actual scattering processes, however, the output state may acquire a phase factor: $|z'\rangle = e^{i\phi}(0,0,1)^T$. Resolving this phase discrepancy represents a significant outstanding challenge worthy of dedicated investigation.

\section{AM and the corresponding atomic motion of non-degenerated phonon at TRIMs}
\label{sec:supp_AM_underT}

In general, the time-reversal operator can be expressed as $\mathcal{T} = U\mathcal{K}$, where $\mathcal{K}$ denotes the complex conjugation operator and $U$ is a finite-dimensional unitary matrix, thus $\mathcal{T}$ is an anti-unitary operator.
For Bosons like phonons, $U$ should be the identity matrix, resulting in $\mathcal{T}=\mathcal{K}$ and $\mathcal{T}^2=1$. 
In this case, the anti-unitary operator has eigenvectors, associated with eigenvalues being arbitrary unitary complex numbers like $e^{i\phi}$.
Thus, for the {non-degenerated} phonon mode with a general form, we have:
\begin{equation}
\begin{aligned}
    \mathcal{T}\boldsymbol{\epsilon}^{\kappa}_{\nu\boldsymbol{q}} & = e^{i\phi} \boldsymbol{\epsilon}^{\kappa}_{\nu\boldsymbol{q}}, \\
    & =e^{i\phi}  \{A^{\kappa}_x e^{i \theta_x}, A^{\kappa}_{y} e^{i\theta_y}, A^{\kappa}_z e^{i \theta_z}\}\\
  &=\{A^{\kappa}_x e^{-i \theta_x}, A^{\kappa}_{y} e^{-i\theta_y}, A^{\kappa}_z e^{-i \theta_z}\}.\\
\end{aligned}
\end{equation}
As a result, 
\begin{equation}
    \begin{aligned}
    \phi  +\theta_\alpha =  -\theta_{\alpha} + 2 n_\alpha \pi, \text{or}\\
     \theta_\alpha = -\phi/2 + n_\alpha  \pi,
\end{aligned}
\end{equation}
where $n_\alpha$ is an integer. Therefore, the phase difference for any of the components relative to the $\kappa$-th atom should be
\begin{equation}
    \Delta_{\alpha\beta} = \theta_\alpha - \theta_\beta = (n_\alpha-n_\beta)\pi. 
    \label{eq:T-delta}
\end{equation}
With Eq.~\ref{eq:T-delta} and Eq.~\ref{eq:relative_phase}, we can write the AM of each atom as
\begin{equation}
    l^\kappa_{\alpha, \nu\boldsymbol{q}}= 2\text{Im}[A^{\kappa}_\beta A^{\kappa}_\gamma e^{i\epsilon_{\alpha(\beta\gamma)} (\theta^\kappa_\beta-\theta^\kappa_\gamma)}],
\end{equation}
Since $\theta_\alpha-\theta_\beta=n\pi$, we have $l^\kappa_{\alpha, \nu\boldsymbol{q}} =0 $. Thus, $\boldsymbol{l}_{\nu\boldsymbol{q}} = \sum_{\kappa}^{N} \boldsymbol{l}^\kappa_{\nu\boldsymbol{q}}$ = 0. In conclusion: 

$\mathcal{T}$ constrains the AM of non-degenerate phonon modes to be zero. From a semi-classical perspective, this corresponds to atomic motions that are either linearly polarized or stationary.

\section{AM and the corresponding atomic motion of non-degenerated phonon at TRIMs under $\mathcal{PT}$}
\label{sec:supp_AM_under_inversion}

Under $\mathcal{PT}$ symmetry, each momentum $\boldsymbol{q}$ remains invariant, as both $\mathcal{P}$ and $\mathcal{T}$ transform $\boldsymbol{q}$ to $-\boldsymbol{q}$. {In this case,} only one-dimensional representations exist, and atoms can be classified into two categories based on their Wyckoff positions.
In class I, atoms $a_{\kappa}$ and $b_{\kappa}$ are related by $\mathcal{P}$. In class II, atoms $o$ are located at the inversion centers. The phonon mode in systems with $\mathcal{PT}$ at any arbitrary $\boldsymbol{q}$ can be expressed in a general form as:
\begin{equation}
    \boldsymbol{\epsilon}_{\nu\boldsymbol{q}}=\{\boldsymbol{\epsilon}^{a_{\kappa}}, \boldsymbol{\epsilon}^{b_{\kappa}}, \boldsymbol{\epsilon}^o\}.
\end{equation}
In phonon systems, $\mathcal{PT}$ serves as an anti-unitary operator with $\mathcal{PT}^2=1$, thereby possessing an eigenvector with an arbitrary eigenvalue $e^{i\phi}$, namely:

\begin{equation}
\begin{aligned}
    \mathcal{PT}\boldsymbol{\epsilon_{\nu\boldsymbol{q}}} &=
    \{-\boldsymbol{\epsilon}^{b_{\kappa}*}, -\boldsymbol{\epsilon}^{a_{\kappa}*}, -\boldsymbol{\epsilon}^{o*}\}\\
  &= e^{i\phi} \{\boldsymbol{\epsilon}^{a_{\kappa}}, \boldsymbol{\epsilon}^{b_{\kappa}}, \boldsymbol{\epsilon}^o\}.
\end{aligned}
\end{equation}
{Hereafter, the amplitude of $a_{\kappa}$ and $b_{\kappa}$ sublattice should be the same, i.e.,}
\begin{equation}
    A^{a_{\kappa}}_{\alpha}=A^{b_{\kappa}}_{\alpha}.
\end{equation}
The relationship of the phases between these sublattices is
\begin{equation}
\label{eq:relative_phase_PT}
    \begin{aligned}
    \boldsymbol{\epsilon}^{b_\kappa*}
    & =e^{i(\phi+(2n+1)\pi)} \boldsymbol{\epsilon}^{a_{\kappa}},\\
    \boldsymbol{\epsilon}^{o*}& =e^{i (\phi+(2n+1)\pi)} \boldsymbol{\epsilon}^{o}.
\end{aligned}
\end{equation}
In comparison to Eq.~\ref{eq:relative_phase_PT}, the relationship between the phases $\theta_\alpha$ for atoms $a_{\kappa}$ and $b_{\kappa}$ in class I can be expressed as
\begin{equation}
    -\theta^{b_\kappa}_{\alpha} = \theta^{a_\kappa}_{\alpha} + \phi +
    (2n_\alpha+1)\pi.
\end{equation}
Thus, the AM for the $a_{\kappa}$ and $b_{\kappa}$ sublattices should be: 
\begin{equation}
    \begin{aligned}
        l^{a\kappa}_{\alpha,\nu\boldsymbol{q}} & = 2\text{Im}[A^{a_{\kappa}}_\beta A^{a_{\kappa}}_\gamma e^{i\epsilon_{\alpha(\beta\gamma)} (\theta^{a_\kappa}_\beta-\theta^{a_\kappa}_\gamma)}],\\
        l^{b\kappa}_{\alpha,\nu\boldsymbol{q}} & = 2\text{Im}[A^{a_\kappa}_\alpha A^{a_\kappa}_\gamma e^{i\epsilon_{\alpha(\beta\gamma)} (-\theta^{a_\kappa}_\beta+\theta^{a_\kappa}_\gamma+2n\pi)}].
\end{aligned}
\end{equation}
$n=n_\alpha-n_\beta$ and it is an integer. Therefore, we obtain
\begin{equation}
    l^{a\kappa}_{\nu\boldsymbol{q}}+l^{b\kappa}_{\nu\boldsymbol{q}} = 0.
\end{equation}
So, phonon AM for a pair of $\mathcal{P}$-related atoms $a_\kappa$ and $b_\kappa$ should be opposite. For atoms $a_{\kappa}$ and $b_{\kappa}$ in class I, they can exhibit linear, circular, or even static motions in the real space. 

For atoms in class II, from Eq.~\ref{eq:relative_phase_PT}, we have 
\begin{equation}
    \begin{aligned}
        -\theta^o_\alpha &= \theta^o_\alpha + \phi + (2n_\alpha+1)\pi,
    \end{aligned}
\end{equation}
thus 
\begin{equation}
    \Delta_{\alpha\beta}=\theta^o_\alpha-\theta^o_\beta=n\pi,
\end{equation}
where $n=n_\alpha-n_\beta$, and it is an integer.
Therefore, phonon AM for atoms in class II is also zero, corresponding to stationary atoms or linear atomic vibrations in real space, while circular motion is forbidden. In conclusion:

$\mathcal{PT}$ enforces zero AM for non-degenerate phonon modes across the entire BZ. For a pair of $\mathcal{P}$-related atoms, their AM values are opposite, corresponding to either opposing circular motion, linear motion, or a static configuration.
Furthermore, atoms located at inversion centers exhibit zero AM, corresponding to phonon modes characterized by stationary atoms or linear atomic vibrations.

\section{AM of excited phonons in high-dimensional IRREPs and the example in which AM and PAM can  be related}
\label{sec:supp_AM_of_emited_phonon}


In the main text, we have proposed that the AM {of the phonon modes} belonging to high-dimensional IRREPs
{can not be determined simultaneously due to the suppositions.} {But it can be}
be determined by the corresponding external stimuli. 
Below, we illustrate this with an example based on the CPRS process (taking the Stoke process as an example) involving the $G$ mode of graphene at the $\boldsymbol{\Gamma}$ point, which has the little group of $D_{6h}$.

\begin{figure}
  \centering
  \includegraphics[width=0.5\textwidth]{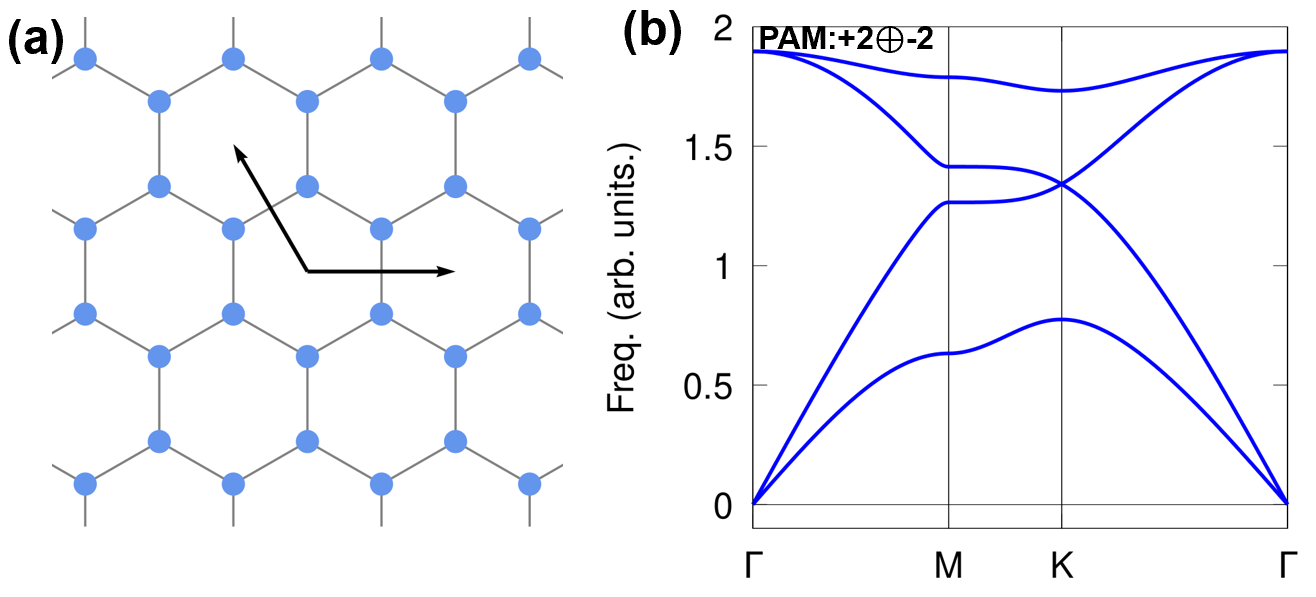}
  \caption{\label{fig:graphene_G}
 (a) Lattice for graphene. 
 (b2) The phonon spectra of graphene.}
\end{figure}

{We first introduce the TB model of graphene.
We set the distance of the nearest-neighbor atoms $a=1$, the lattice constant of graphene should to be $\sqrt{3}$, the lattice vectors are $\boldsymbol{a}_1=(\sqrt{3},0)$, $\boldsymbol{a_2}=(-\frac{\sqrt{3}}{2}, \frac{3}{2})$, and the corresponding reciprocal lattice vectors are $b_1=2\pi(\frac{1}{\sqrt{3}}, \frac{1}{3})$, $b_2=2\pi(0,\frac{2}{3})$. There are two sublattices in the primitive unit cell, located at $s_1=(\frac{\sqrt{3}}{2}, -\frac{1}{2})$, $s_2=(\frac{\sqrt{3}}{2}, \frac{1}{2})$, and the masses of them are set to be 1.
Here, we only take the nearest-neighbor interaction.
The longitudinal spring constant $L$ is set to be 1, and the transverse spring constant $T$ is set to be 0.2, respectively. The phonon spectrum is shown in Figure~\ref{fig:graphene_G} (b).}

By neglecting the $z$-component degree of freedom, the point group reduces to $C_{6v}$. In the CPRS process, vertical mirror symmetry is broken because CPL is not an eigenstate of $\mathcal{T}$ nor $\mathcal{M}$, and CPL acts as a perturbation breaking these symmetries. Therefore, we focus on the representation matrix of $C_{6}$. {The atoms are not located at the $C_{6}-$symmetric Wyckoff positions, and $D(C_6)$ has the non-diagonalized form} :
\begin{equation}
\label{eq:C6_graphene}
    {D(C_6)= 
    \begin{pmatrix}
        0 & C_6 \\
        C_6 & 0
    \end{pmatrix}.}
\end{equation}
The $G$  mode is composed of the phonon modes with PAM=$\pm 2$, i.e., the $E_2$ {IRREP} in Table~\ref{tab:graphene_gamma}. For the relative phase between the two sublattices is $\pi$, $G$ is the optical mode.
In the equilibrium state without CPRS, both the $\mathcal{T}$ and the vertical mirror enforce the degeneracy of phonon modes with PAM=$\pm2$. Consequently, phonon AM in this space remains undetermined.

While under the CPRS, only the phonon mode with PAM=$-2$ is active and emitted in the $\sigma^{-}/\sigma^{+}$ process, according to Figure~\ref{fig:Raman_antiRaman} (b).
In this specific case, determined by the unique symmetry and the occupied Wyckoff positions, the PAM and AM exhibit a one-to-one correspondence, as shown in Table~\ref{tab:graphene_gamma}. Thus, the phonon emitted in the $\sigma^{-}/\sigma^{+}$ process has AM of $+1$. However, in more general cases, this relationship may not hold, and the AM of the emitted phonon can be nonzero, as discussed in the main text.

Although CPL breaks mirror symmetry as a perturbation, it typically does not alter phonon dispersion or frequency due to the weak direct phonon-light scattering cross-section. Instead, it acts as a filter, fixing the coefficients for the superposition of phonon modes with PAM =$\pm2$. 

\section{The AM and atomic motion for the phonon mode detected by CPRS experiments.}
\label{sec:supp_atom_motion_exp}

\begin{figure}
  \centering
\includegraphics[width=1.0\textwidth]{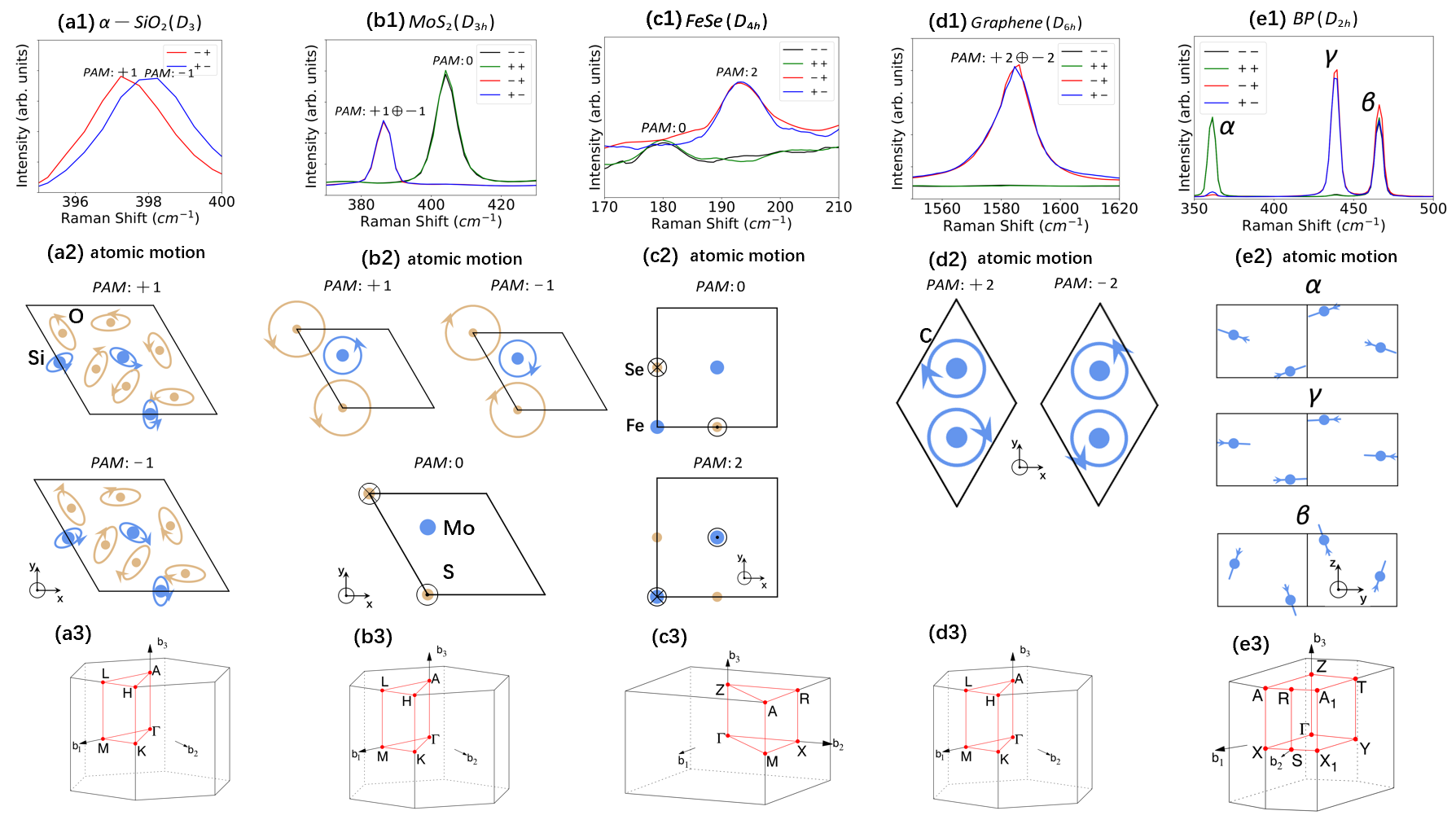}
\caption{\label{fig:supp_raman_atom_motion}
(a1)-(e1) The Raman spectra for $\alpha-$SiO$_2$, MoS$_2$, FeSe, graphene, and BP.
(a2)-(e2) The atomic motion corresponding to the phonon modes excited by the CPRS process. 
(c3)-(e3) The labels of the high-symmetry points in BZ for the corresponding materials.} 
\end{figure}

To give experimental validation of our theoretical proposal, we performed the CPRS for five materials. 
Figure~\ref{fig:supp_raman_atom_motion} (a1)-(e1) show the Raman spectra for $\alpha-$SiO$_2$, MoS$_2$, FeSe, graphene, and BP, which are represented here for reference. Figure~\ref{fig:supp_raman_atom_motion} (a2)-(e2) show the corresponding atomic motion from DFT for the phonon modes detected by CPRS for these materials, which is consistent with the results in the main text.
The labels of the high-symmetry points in BZ are shown in Figure~\ref{fig:supp_raman_atom_motion} (a3)-(e3).
For graphene, $C_6$ eigenstates and the AM of the emitted phonon are shown in Table~\ref{tab:graphene_gamma}.

\clearpage

\begingroup
\setlength{\tabcolsep}{12pt} 
\renewcommand{\arraystretch}{1.5} 
\begin{table}[]
\centering
\caption{\label{tab:C6_1} The atomic position of the $C_{6_1}$ screw chain in Cartesian coordinate}
\begin{tabular}{c|c}
\hline
\hline
Index    & Position (Arb. Units)    \\ \hline                  
$\kappa_1$ &    ( 0.1000,   0.5196,    0.0000) \\ \hline 
$\kappa_2$ &    (-0.4000,   0.3464,    1.0000) \\ \hline 
$\kappa_3$ &    (-0.5000,  -0.1732,    2.0000) \\ \hline
$\kappa_4$ &    (-0.1000,  -0.5196,    3.0000) \\ \hline 
$\kappa_5$ &    ( 0.4000,  -0.3464,    4.0000) \\ \hline 
$\kappa_6$ &    ( 0.5000,   0.1732,    5.0000) \\ \hline
\end{tabular}
\end{table}
\endgroup

\begingroup
\setlength{\tabcolsep}{7pt} 
\renewcommand{\arraystretch}{1.1} 
\begin{table}[h]
\centering
\caption{\label{tab:supp_PAM_AM_C2} The relationship between AM and PAM under $C_2$ symmetry and specific occupied Wyckoff positions (WP)}
\begin{tabular}{c|c|c}
\hline
\hline
Site symmetry of WP        & PAM                              & AM   \\ \hline
$C_2$                                             &  0    &    0        \\ \hline
\end{tabular}
\end{table}
\endgroup

\begingroup
\setlength{\tabcolsep}{7pt} 
\renewcommand{\arraystretch}{1.1} 
\begin{table}[h]
\centering
\caption{\label{tab:supp_PAM_AM_C3} The relationship between AM and PAM under $C_3$ symmetry and specific occupied Wyckoff positions (WP)}
\begin{tabular}{c|c|c}
\hline
\hline
Site symmetry of WP        & PAM                           &   AM   \\ \hline
       ~                                           & $+1$  &   $+1$      \\ \cline{2-3}
$C_3$                                              & $-1$  &  $-1$       \\ \cline{2-3}
       ~                                           &  0    &    0        \\ \hline
\end{tabular}
\end{table}
\endgroup

\begingroup
\setlength{\tabcolsep}{7pt} 
\renewcommand{\arraystretch}{1.1} 
\begin{table}[h]
\centering
\caption{\label{tab:supp_PAM_AM_C4} The relationship between AM and PAM under $C_4$ symmetry and specific occupied Wyckoff positions (WP).}
\begin{tabular}{c|c|c}
\hline
\hline
Site symmetry of WP        & PAM                              & AM    \\ \hline
       ~                                           & $+1$  &   $+1$      \\ \cline{2-3}
       ~    $C_4$                                  & $-1$  &  $-1$       \\ \cline{2-3}
       ~                                           &  0    &    0        \\ \hline
       ~    $C_2$                                  & $2$   &  0       \\ \hline       
\end{tabular}
\end{table}
\endgroup

\begingroup
\setlength{\tabcolsep}{7pt} 
\renewcommand{\arraystretch}{1.1} 
\begin{table}[h]
\centering
\caption{\label{tab:supp_PAM_AM_C6} The relationship between AM and PAM under $C_6$ symmetry and specific occupied Wyckoff positions (WP).}
\begin{tabular}{c|c|c}
\hline
\hline
Site symmetry of WP         & PAM                              & AM     \\ \hline
       ~                                           & $+1$  &   $+1$      \\ \cline{2-3}
       ~    $C_6$                                  & $-1$  &  $-1$       \\ \cline{2-3}
       ~                                           &  0    &    0        \\ \hline
~                                                  & $+1$   &  $+1$ \\ \cline{2-3}
~                                                  & $-1$   &  $-1$ \\ \cline{2-3} 
       ~    $C_3$                                  & $+2$   &  $-1$ \\ \cline{2-3}
~                                                  & $-2$   &  $+1$ \\ \cline{2-3}
~                                                  & $0, 3$ &  $0$  \\ \hline 
\end{tabular}
\end{table}
\endgroup

\begingroup
\setlength{\tabcolsep}{10pt} 
\renewcommand{\arraystretch}{1.5} 
\begin{table}[h]
\caption{\label{tab:graphene_gamma} The eigenvectors, eigenvalues (expressed as PAM), {AM}, and irreducible representations~(IRREPS) under $C_6$ operator, which corresponds to the little group at momentum  $\boldsymbol{\Gamma}$ for graphene lattice.}
\centering
\begin{tabular}{c|c|c|c|c}
\hline
\hline
Label    & State                                        & PAM      &   AM       & IRREPs \\ \hline
$\epsilon_{1\boldsymbol{\Gamma}}$ & $\frac{1}{2} (1,i,-1,-i)^T$ & $-2$     &   $+1$     & $E_2$ \\ \hline
$\epsilon_{2\boldsymbol{\Gamma}}$ & $\frac{1}{2} (1,-i,-1,i)^T$ & $+2$     &   $-1$     & $E_2$    \\ \hline
$\epsilon_{3\boldsymbol{\Gamma}}$ & $\frac{1}{2} (1,-i,1,-i)^T$ & $-1$     &   $-1$     & $E_1$  \\ \hline
$\epsilon_{4\boldsymbol{\Gamma}}$ & $\frac{1}{2} (1,i,1,i)^T$   & $+1$     &   $+1$     & $E_1$  \\ \hline
\end{tabular}
\label{tab:hBN}
\end{table}
\endgroup
